\documentclass[usegraphicx,usenatbib]{mn2e}

\title[Galaxy properties in different environments up to $z\sim3$]{Galaxy properties in different environments up to $z\sim3$ in the GOODS NICMOS Survey.}

\author[Gr\" utzbauch et al.]{Ruth Gr\"utzbauch$^{1}$\thanks{email: ruth.grutzbauch@nottingham.ac.uk}, Robert W. Chuter$^{1}$,
Christopher J. Conselice$^{1}$, \newauthor Amanda E. Bauer$^{1}$, Asa F. L. Bluck$^{1}$, Fernando Buitrago$^{1}$, Alice Mortlock$^{1}$ \\
$^{1}$ School of Physics and Astronomy, University of Nottingham, UK
}

\begin{document}

\date{Accepted 2010 November 22.  Received 2010 November 5; in original form 2010 June 18}

\maketitle

\begin{abstract}

We study the relationship between galaxy colour, stellar mass, and local galaxy density in a deep near-infrared imaging survey up to a redshift of $z\sim3$ using the GOODS NICMOS Survey (GNS). The GNS is a deep near-infrared Hubble Space Telescope survey imaging a total of 45 arcmin$^2$ of the GOODS fields, reaching a stellar mass completeness limit of $M_\ast = 10^{9.5}~M_\odot$ at $z=3$. Using this data we measure galaxy local densities based on galaxy counts within a fixed aperture, as well as the distance to the 3$^{rd}$, 5$^{th}$ and 7$^{th}$ nearest neighbour. We compare the average rest-frame $(U-B)$ colour and fraction of blue galaxies in different local densities and at different stellar masses. We find a strong correlation between colour and stellar mass at all redshifts up to $z\sim3$. Massive red galaxies are already in place at $z\sim3$ at the expected location of the red-sequence in the colour-magnitude diagram, although they are star forming.
We do not find a strong correlation between colour and local density, however, there may be evidence that the highest overdensities are populated by a higher fraction of blue galaxies than average or underdense areas. This could indicating that the colour-density relation at high redshift is reversed with respect to lower redshifts ($z<1$), where higher densities are found to have lower blue fractions. Our data suggests that the possible higher blue fraction at extreme overdensities might be due to a lack of {\it massive} red galaxies at the highest local densities. 

\end{abstract}


\section{Introduction}

It is now well-established that the most massive galaxies formed very early in the history of the universe and are mainly in place in their present form by a redshift of $z \sim 1$ \citep[see e.g.][]{Mad96,Cim04,Jun05,Con07}, when the universe was only about half its present age. These massive galaxies largely stopped forming stars at redshifts $z > 1$, and their stellar populations subsequently for the most part evolved and reddened passively. The bulk of the star formation activity in the universe shifts to lower and lower mass galaxies as cosmic time proceeds \citep[see e.g.][]{Bau05,Bun06}. 
This mass dependent shift in star formation activity might be the basis of the observed bimodality in colour-magnitude space, where we see a population of old, passive, red galaxies following a relatively tight correlation between colour and magnitude, the so-called red sequence, and a population of young, starforming, blue galaxies, which occupy a rather diffuse area in colour-magnitude space, known as the blue cloud. The presence of a strong correlation between galaxy colour and stellar mass is supported by numerous observational studies in the local universe \citep[see e.g.][]{Kau03a,Kau03b} and up to intermediate redshifts of $z\sim1$ \citep[see e.g.][]{Bun06,Gru10}.

However, the colour-magnitude relation (and related quantities) not only depends on mass, but also on a galaxy's environment. The preference of red galaxies for denser local environments, first noticed by \citet{Oem74}, and confirmed by \citet{Dre80}, is now well-studied in the local universe \citep[see e.g.][]{Kau04,vdW08}. In the early universe, however, the evidence is controversial. While some studies find that the colour-density relation at $z\sim1$ is mainly due to a bias in stellar mass selection and only persists for low-mass galaxies \citep[see e.g.][]{Tas09,Iov10,Gru10}, others argue that a strong colour-density relation is already in place at $z\sim1.4$ \citep[see e.g.][]{Coo06,Ger07} even at fixed stellar mass \citep{Coo10}.

We know that the bulk of star formation occurs in the early universe with half of the currently existing stellar mass already in place by $z=1$ \citep{Bri00,Dro04,Per08}. However, it is still unclear at which redshift the star formation rate density peaks and how it evolves at higher $z$. It is also unknown where most of the star formation happens in the early universe. Does the local SFR-density relation turn around by $z \sim 1$, as reported by some authors \citep{Elb07,Coo08}, or is the local morphology-density relation already present early in the history of the universe \citep[e.g.][]{Pan09}?

We also know that massive red galaxies exist at very high redshifts, but what are their typical environments? Are they already found preferentially at higher local densities like most red galaxies in the local universe \citep[see e.g.][]{Kau04}? Can we witness their transformation from heavily star-forming to passively evolving galaxies, i.e. the build-up of the red sequence in the colour-magnitude relation? If so, when and where does it happen? Does the red sequence grow ``inside-out'', from the densest to the least dense regions, or is it the stellar mass that predominantly determines the evolutionary path of a galaxy?

In this study we intend to answer some of the above questions by investigating the relationship between galaxy colours, stellar mass and local densities in the early universe at $z>1$. Recent high-$z$ studies of the environmental influence on galaxy evolution have focused on investigating the presence of bright neighbours and the incidence of major merging \citep[see e.g.][]{Con03,Con08b,Blu09}. To study the influence of local environment on a wider range of scales, including also other environment-related effects like minor merging or galaxy harassment, the depth of the survey is critical. Additionally, we want to disentangle the influence of environment and stellar mass, with a sample that is volume limited with the same stellar mass completeness limit at all colours and redshifts.

The sample we use in this study is drawn from the GOODS NICMOS Survey (GNS), a large Hubble Space Telescope survey of unprecedented depth, reaching galaxies $\sim$200 times less luminous than available from ground based near-infrared surveys. The GNS samples over two orders of magnitude in stellar mass with a stellar mass completeness limit of $M_\ast \sim 10^{9.5}~M_\odot$ at $z\sim3$. 

The paper is structured as follows: Section~2 presents the GNS survey and data, the sample selection and measurement of colours and stellar masses. Section~3 describes how we quantify galaxy environment, and Section~4 presents the results: the colour-magnitude relation as well as rest-frame colour and blue fraction as a function of stellar mass and local density. Section~5 summarises our conclusions. We use standard cosmological parameters (H$_{0}=70$ km~s$^{-1}$ Mpc$^{-1}$, $\Omega _{m}=0.3$, and $\Omega _{\Lambda }=0.7$) and the $AB$ magnitude system throughout the paper.

\section{The sample}

\subsection{The GOODS NICMOS Survey}

The GOODS NICMOS Survey (GNS) is a 180 orbit Hubble Space Telescope survey consisting of 60 pointings with the NICMOS-3 near-infrared camera. Each pointing is centred on a massive galaxy ($M_\ast > 10^{11}~M_\odot$) in the redshift range $1.7 < z < 3$, selected by their optical-to-infrared colours \citep[see][]{Con10} from the GOODS (Great Observatories Origins Deep Survey) fields \citep{Dic03}. GOODS is designed to compile deep observational data over two fields of about 320 arcmin$^2$ and a wide range of the electromagnetic spectrum, ranging from Chandra and XMM-Newton X-ray data, over Hubble Space Telescope (HST-ACS) optical imaging to Spitzer IRAC and MIPS observations in the infrared. The extensive imaging data is complemented by spectroscopic surveys using the Very Large Telescope (VLT) spectrographs FORS2 and VIMOS \citep{Van05,Van06,Van08,Pop09} for the GOODS South field, and DEIMOS at the Keck Observatory to cover GOODS North \citep{Bar08}. The UV-to-infrared photometric and spectroscopic data available in the GOODS South field are compiled by the FIREWORKS project \citep{Wuy08}. In this study we use the optical HST-ACS imaging data \citep{Gia04} and spectroscopic redshifts from all available spectroscopic surveys.

The positions of the 60 GNS pointings were optimised to contain as many massive galaxies as possible and are partly overlapping, covering a total area of about 45 arcmin$^2$ \citep{Con10}. The field of view of the NICMOS-3 camera is $51.2 \times 51.2$ arcsec$^2$ with a pixel scale of 0.203 arcsec/pixel. The single exposures were dithered to obtain a higher resolution of $\sim$0.1 arcsec/pixel. The PSF has a width of $\sim$0.3 arcsec FWHM. 
The data reduction is described in detail by Magee, Bouwens \& Illingworth (2007). A full description of the survey as well as the target selection will be given in \citet{Con10}. The GNS is also described in \citet{Bui08}, \citet{Blu09}, \citet{Blu10} and Bauer et al. (submitted).

\subsection{Source detection and catalogue cross-matching}\label{sourcedetection}

Sources are detected in the $H$-band images using {\tt SExtractor} \citep{Ber96}. We use a 2 sigma threshold above the background noise and a minimum number of 3 adjacent pixels with values above this threshold. Magnitudes are measured using the {\tt MAG\_AUTO} output parameter, which measures the flux in a Kron-like elliptical aperture, where the aperture is not fixed but depends on the light distribution of the galaxy. The flux within the aperture is then converted to magnitudes using the magnitude zeropoint of 25.17.
The limiting magnitude reached at 5$\sigma$ is $H_{AB}$ = 26.8. This is significantly deeper than ground based near-infrared imaging of the GOODS fields done with ISAAC on the VLT, reaching a 5$\sigma$ depth of $H_{AB} =24.5$ \citep{Ret10}.

\begin{figure*}
\includegraphics[angle=270, width=0.72\textwidth]{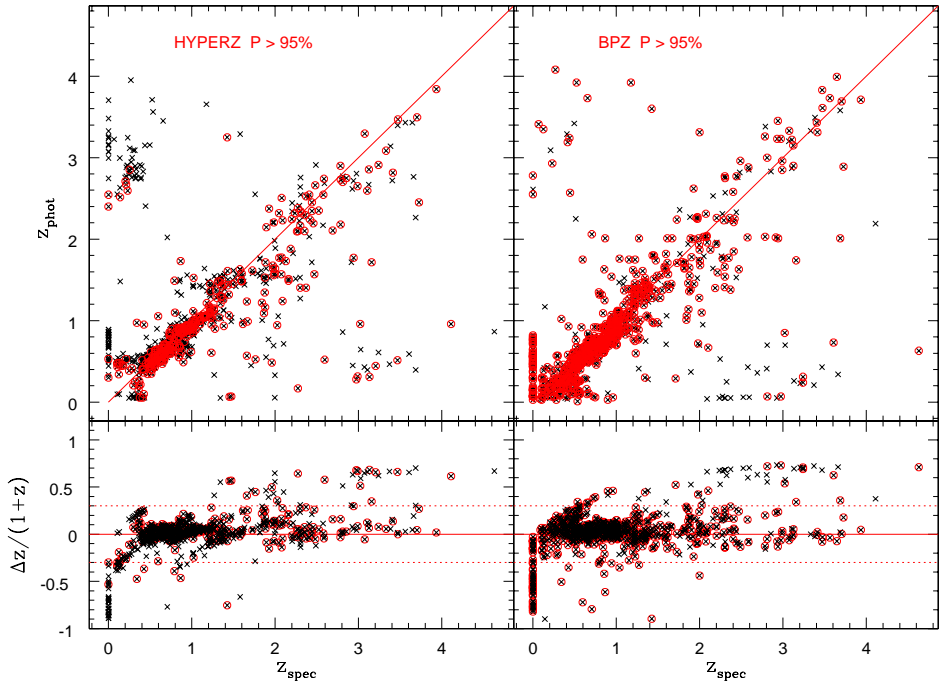}
\caption{Reliability of photometric redshifts: {\it Top panels:} photometric vs. spectroscopic redshifts using HYPERZ (left) and BPZ (right). Photometric redshifts with a high probability of $P > 95\%$ are circled in red. {\it Bottom panels:} $\Delta z/(1+z)$ dependent on redshift. Black symbols show all redshifts, red symbols high probability redshifts only. The dashed lines show the limit for catastrophic outliers at $|\Delta z/(1+z)| > 0.3$.}
\label{fig1}
\end{figure*}

Stars, and obvious spurious detections, were removed from the catalogue, yielding a total of 8298 galaxies in the final full catalogue.
To obtain photometric redshifts, stellar masses and rest-frame colours, the NICMOS $H$-band sources were matched to a catalogue of optical sources in the GOODS ACS fields. Photometry in the $B$, $V$, $i$ and $z$ bands is available for sources down to a limiting magnitude of $B_{AB} \sim 28.2$. The matching is done within a radius of $2 ^{\prime\prime}$, however the mean separation between optical and $H$-band coordinates is much better with $\sim 0.28 \pm 0.4 ^{\prime\prime}$, roughly corresponding to the NICMOS resolution. We identify 1076 complete drop-outs in our $H$-band selected catalogue, i.e., galaxies that are detected in $H$-band, but have no counterpart in any of the ACS bands. This corresponds to a drop-out rate of 13\%. The nature of these sources and a full analysis on the likelihood of some of them being $z\sim7$ galaxies is presented in \citet{Bou10}.

\subsection{Photometric redshifts}\label{photoz}

The photometric redshifts were obtained by fitting template spectra to the $BVizH$ photometric data points. The degeneracy in color-redshift space is problematic, especially when few filters are available. To cope with this effect we used two different approaches: the standard $\chi ^2$ minimisation procedure, using HYPERZ \citep{Bol00} and the Bayesian approach using BPZ \citep{Ben00}. 
The synthetic spectra used by HYPERZ are constructed with the Bruzual \& Charlot evolutionary code \citep{Bru93} representing roughly the different morphological types of galaxies found in the local universe. We use 5 template spectra corresponding to the spectral types of E, Sa, Sc and Im as well as a single burst scenario. The reddening law is taken from \citet{Cal00}. The photo-z code then computes the most likely redshift solution in the parameter space of age, metallicity and reddening. The best fit redshift and corresponding probability are then output together with the best fit parameters of spectral type, age, metallicity, $A_V$ and secondary solutions.
 
The Bayesian approach uses a similar template fitting method, apart from using empirical rather than synthetic template SEDs. The main difference though is that it relies not only on the maximum likelihood of the redshift solution in the parameter space as described above. Instead it uses additional empirical information about the likelihood of a certain combination of parameters, also called prior information. The redshift solution with the maximum likelihood is determined after weighting the probability of each solution by the additional probability determined from the prior information. In our case the prior is the distribution of magnitudes for the different morphological types as a function of redshift, obtained from the Hubble Deep Field North (HDF-N) Data \citep{Ben00}. In short, the code not only determines the best fit redshift and spectral type, but takes into account how likely it is to find a galaxy of that spectral type and magnitude at the given redshift.

To assess the reliability of our photometric redshifts we compare them to available spectroscopic redshifts in the GOODS fields. Spectroscopic redshifts of sources in the GOODS-N field were compiled by \citet{Bar08}, whereas the GOODS-S field spectroscopic redshifts are taken from the FIREWORKS compilation \citep{Wuy08}. We matched the two catalogues to our photometric catalogue with a matching radius of $2 ^{\prime\prime}$, obtaining 537 spectroscopic redshifts for our sources in GOODS-N and 369 in GOODS-S, adding up to 906 redshifts in total. The mean separation between photometric and spectroscopic sources is $0.41 \pm 0.06$ arcsec in the GOODS-N field and $0.13 \pm 0.05$ arcsec in the GOODS-S field. The reliability of photometric redshift measures is usually defined as $\Delta z/(1+z) \equiv (z_{spec}- z_{phot})/(1+z_{spec})$. In the following we compare the median offset from the one-to-one relation between photometric and spectroscopic redshifts, $\langle \Delta z/(1+z) \rangle$, and the RMS scatter around this relation, $\sigma_{\Delta z/(1+z)}$. 

We find good agreement between photometric and spectroscopic redshifts for both codes, however, HYPERZ is slightly better at identifying outliers by assigning low probabilities to their redshifts. The comparison between photometric and spectroscopic redshifts is shown in Figure~\ref{fig1}. The photometric redshifts of HYPERZ (left) and BPZ (right) are plotted against the spectroscopic redshifts available in the literature. High probability $z_{phot}$ are circled in red. 
The lower panel shows the $\Delta z/(1+z)$ dependence on redshift $z_{spec}$: there is no clear trend with redshift, with perhaps a slight trend to underestimate $z$ at high redshifts.
We obtain the following results for HYPERZ: when considering only high probability photometric redshifts ($P > 95\%$), the median offset is $\langle \Delta z/(1+z) \rangle = 0.033$, with a scatter of $\sigma_{\Delta z /(1+z)} = 0.045$ (356 out of 906 galaxies with $P > 95\%$). Using all redshifts regardless of their probability, the median offset decreases to $\langle \Delta z/(1+z) \rangle = 0.011$, while the scatter rises slightly to $\sigma_{\Delta z /(1+z)} = 0.061$. BPZ gives comparable offsets, but slightly higher scatters: $\langle \Delta z/(1+z) \rangle$ = 0.026 with a scatter of $\sigma_{\Delta z /(1+z)} = 0.058$ for high probability redshifts (792 out of 906) and $\langle \Delta z/(1+z) \rangle = 0.030$ and $\sigma_{\Delta z /(1+z)} = 0.064$ for all galaxies. BPZ identifies more redshifts with a high probability than HYPERZ, however, they are not necessarily better than the low probability redshifts, as shown by the comparable scatter. Overall, the performance of HYPERZ is slightly better than that of BPZ.

\begin{figure}
\includegraphics[width=0.48\textwidth]{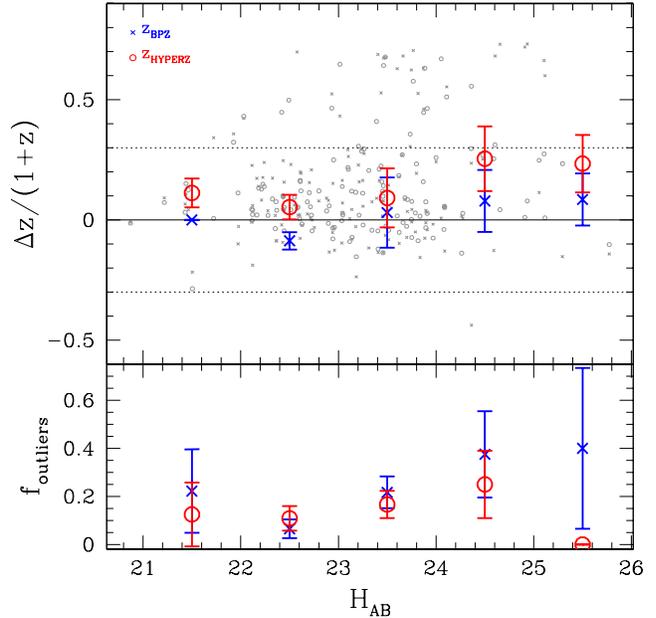}
\caption{Dependence of $\Delta z/(1+z)$ on magnitude.{\it Top panel:} $\Delta z/(1+z)$ vs. $H$-band magnitude for HYPERZ (red circles) and BPZ (blue crosses). Median values in each magnitude bin (bin size 1 mag) are plotted as big symbols with errorbars representing the scatter in each bin. The dashed lines show the limit for catastrophic outliers at $|\Delta z/(1+z)| > 0.3$}. {\it Bottom panel:} fraction of catastrophic outliers $|\Delta z/(1+z)| > 0.3$ as a function of $H$-band magnitude. Only galaxies in the redshift range $1.5 < z < 3$ are used in this plot.
\label{fig2}
\end{figure}

Figure~\ref{fig2} shows the dependence of $\Delta z/(1+z)$ on $H$-band magnitude. HYPERZ and BPZ results are plotted as circles and crosses respectively. Only galaxies in the redshift range of $1.5 \leq z \leq 3$ are shown in this plot. The median offset and RMS scatter are computed in each magnitude bin and are plotted as big symbols (with corresponding errorbars) in red for the HYPERZ output and in blue for the BPZ output. The figure shows the slightly better performance of HYPERZ, which is also visible in the fraction of catastrophic outliers with $|\Delta z/(1+z)| > 0.3$ (bottom panel of Figure~\ref{fig2}). The redshift error is stable up to faint magnitudes of $H_{AB} \sim 24$, as is the fraction of outliers.

In the following we use the HYPERZ photometric redshifts, regardless of their probability. We use all probability redshifts since this gives us a much larger sample of galaxies with only a slightly higher scatter. 
We now investigate the performance of HYPERZ at different redshifts, at low redshift ($z<1.5$) and in the redshift range of $1.5 \leq z \leq 3$, which is the redshift range of the galaxy sample we use in this study. These are the values that are used in the Monte Carlo Simulations described in Section~\ref{MC} to account for the photometric redshift errors.
For the high redshift sample we obtain an average offset $\langle \Delta z/(1+z) \rangle = 0.06$ and a RMS of $\sigma_{\Delta z /(1+z)} = 0.10$, with a fraction of catastrophic outliers of $20\%$. As above catastrophic outliers are defined as galaxies with $|\Delta z/(1+z)| > 0.3$, which corresponds to $\sim$ 3 times the RMS scatter.
Galaxies below $z=1.5$ show a slightly lower, but still comparable scatter of $\sigma_{\Delta z /(1+z)} = 0.08$, however the outlier fraction decreases dramatically to only $\sim 2\%$.

 The comparison of our results with photometric redshifts already available for the brighter part of our sample shows that our photometric redshifts are of comparable quality. Photometric redshifts taken from the FIREWORKS compilation have a median offset of $\langle \Delta z/(1+z) \rangle = -0.037$ and a RMS scatter of $\sigma_{\Delta z /(1+z)} \sim 0.028$, compared to $\langle \Delta z/(1+z) \rangle = 0.011$ and $\sigma_{\Delta z /(1+z)} = 0.061$ for our full sample. 

\subsection{Stellar masses and rest frame $(U-B)$ colours}\label{stellar mass and colours}

The stellar masses and rest-frame colours we use are measured by multicolour stellar population fitting techniques, based on the same catalogue used for the photometric redshift measurements. Spectroscopic redshifts are used if available. A large set of synthetic spectral energy distributions (SEDs) is constructed from the stellar population models of \citet[][BC03]{Bru03}, assuming a Salpeter initial mass function. The star formation history is characterised by an exponentially declining model with various ages, metallicities and dust extinctions. These models are parametrised by an age of the onset of star formation, and by an e-folding time such that

\begin{equation}
{\rm SFR (t)} \sim\, {\rm SFR_{0}} \times {\rm e}^{-\frac{t}{\tau}}
\end{equation}

\noindent where the values of $\tau$ are randomly selected from a range between 0.01 and 10 Gyr, while the age of the onset of star formation ranges from 0 to 10 Gyr. The metallicity ranges from 0.0001 to 0.05 (BC03), and the dust content is parametrised by $\tau_{\rm v}$, the effective V-band optical depth for which we use values $\tau_{\rm v} = 0.0, 0.5, 1, 2$. 

The model SEDs are then fit to the observed photometric datapoints of each galaxy using a Bayesian approach. For each galaxy a likelihood distribution for the stellar mass, age and absolute magnitude at all star formation histories described above is computed. We chose to compute rest-frame $(U-B)$ colours, since the wavelength range of the $U$ and $B$ bands is covered best by the observed optical and H bands.
The peak of the likelihood distribution is then adopted as the galaxy's stellar mass and absolute $U$ and $B$-band magnitude, while the uncertainty of these values is given by the width of the distribution. The rest-frame $(U-B)$ colours are obtained by subtracting the two absolute magnitudes.
While parameters such as age, e-folding time and metallicity are not accurately fit due to various degeneracies, the stellar masses and colours are robust. From the width of the probability distributions we determine typical errors for our stellar masses of 0.2 dex. There are additional uncertainties from the choice of the IMF and due to photometric errors, resulting in a total random error of our stellar masses of $\sim$0.3 dex, roughly a factor of two.  

We do not probe with our filters redder than roughly the rest-frame $B$-band at $z\sim3$ in our measurements of stellar masses.   The reason for this is similar to our reasons for not using longer wavelength data for the photometric redshifts - essentially there is no data redder than the $H-$band which has the same fidelity and depth as the $BVizH$ bands we use in this paper.  The K-band data available from the ground is no where near as deep as the $H-$band NICMOS data.  While we have IRAC data for our sources, we do not use these data due to the PSF issues and contamination from neighbouring galaxies.    Furthermore, the rest-frame $B$-band gives us a good anchor for measuring stellar masses, as is shown by e.g., \citet{BdJ01} at lower redshifts, and \citet{Bun06} for higher redshifts.

There is also a question as to whether or not our stellar masses are overestimated by the choice of stellar population models used.  It has been argued by \citet{Mar06}, among others, that the \citet{Bru03} models we use here can result in stellar masses that are too high by a factor of a few due to how thermal-pulsating AGB stars are treated. While we consider random uncertainties of a factor of two in our stellar masses, it is worth investigating whether or not our stellar masses could suffer from a systematic bias.
\citet{Mar06} have determined that galaxy stellar masses computed with an improved treatment of TP-AGB stars are roughly 50-60\% lower. However, the effect of TP-AGB stars is less important at the rest-frame wavelengths we probe than it is at longer wavelengths, especially in the rest-frame IR. Our sample is selected in the $H$-band, which corresponds roughly to rest-frame $R$-band at $z\sim1.5$ and approximately rest-frame $B$-band at $z\sim3$, where the effects of TP-AGB stars are negligible, as shown in previous work using similar data and the same stellar mass code \citep{Con07}. In this earlier study the effect of TP-AGB stars was tested on a sample of massive galaxies by using the newer Bruzual and Charlot (in preparation) models, which include a new TP-AGB star prescription. From this we find that stellar masses are on average 0.07 dex smaller than the ones calculated with the BC03 models.  This systematic error is however much smaller than both the random error we assume (0.3 dex), and the cosmic variance uncertainties, and thus we conclude it is not a significant factor within our analysis.

\subsection{Colour and completeness limits of red and blue galaxies}\label{completeness-colour}

To investigate the different behaviour of red and blue galaxies we split the sample into red and blue populations by their location in the colour-magnitude diagram.  
Due to the low number of galaxies in our sample and the lack of accurate spectroscopic redshifts for the majority of objects, we do not attempt to fit the red sequence of the colour-magnitude relation in this study. Instead we evolve the red sequence found at lower redshift back in time as expected from the passive colour evolution of an old stellar population. This procedure is valid due to the high formation redshift ($z>3$) of red sequence galaxies in stellar population models. After the star formation has stopped, the stellar populations of these galaxies are supposedly ageing (=reddening) quickly and subsequently evolving passively \citep{vDF01}. We take the red sequence of galaxies in the DEEP2 redshift survey at $z\sim1$ \citep{Wil06}, converted to AB magnitudes:

\begin{equation}
(U-B) = -0.032~(M_B+21.52)+1.284
\end{equation}

\noindent and evolve it back with redshift according to \citet{vDF01}. The separation between the narrow red sequence and the diffuse blue cloud was found to be present up to at least $z\sim1.4$ \citep{Wil06}, with a separation between the two populations of 0.25 magnitudes. We adopt this colour limit of 0.25 mag blueward of the red sequence to separate between red and blue galaxies.

\begin{figure}
\includegraphics[width=0.48\textwidth]{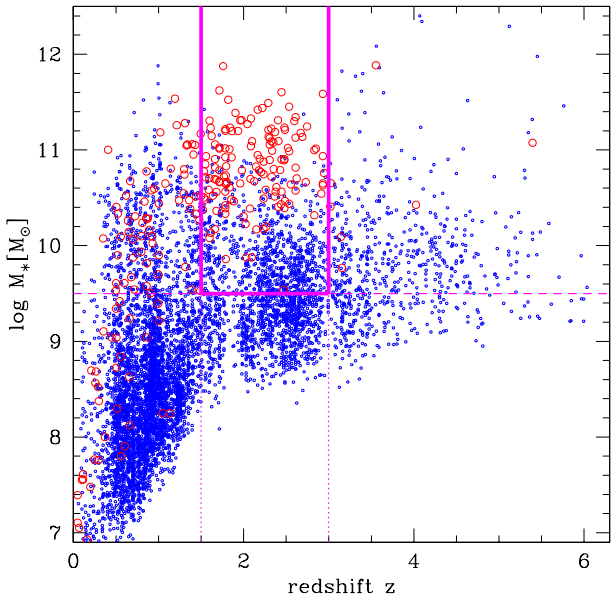}
\caption{Distribution of stellar masses ($\log M_\ast$) with redshift $z$ for red galaxies (red big circles) and blue galaxies (blue small circles). The adopted completeness limit of $\log M_\ast = 9.5 M_\odot$ (at $z=3$) is shown as dashed line. The limits of the redshift range used in this study are marked with dotted lines. The resulting window in redshift and stellar mass containing the galaxy sample we use in the following is given by the thick solid lines.\label{fig3}}
\end{figure}

We compute the expected completeness limits for red and blue galaxies from the 5$\sigma$ magnitude limit of the GNS (see Section~\ref{sourcedetection}) and the mass-to-light ratios (M/L) of simple stellar populations (SSP) assuming a galaxy as old as the universe at each redshift of interest who formed all of their stars in one initial burst.   For this purpose M/L at the rest-frame wavelength corresponding to the observed $H$-band at $z\sim3$ is determined for different SSPs using the models of \citet{Wor94}. To obtain the limiting stellar mass, the M/L is then multiplied with the luminosity corresponding to a $z=3$ galaxy at the detection limit of $H_{AB}$ = 26.8. To compute the completeness limit for red galaxies we use an evolved SSP with an age of 2 Gyr, corresponding to the age of the universe at $z\sim3$. A young SSP with an average age of 1 Gyr is used to obtain the limiting stellar mass for the blue galaxy population. We obtain limiting stellar masses of $M_\ast = 10^{8.9} M_\odot$ for blue galaxies and $M_\ast = 10^{9.2} M_\odot$ for red galaxies respectively at $z = 3$.

In Figure~\ref{fig3} we plot stellar mass versus redshift $z$ for blue and red galaxies, as defined above. The dashed magenta line shows the completeness limit of $\log M_\ast = 9.5$ at the upper redshift limit of $z=3$ (dotted line) we adopted in this study. The completeness of blue galaxies at a given stellar mass is higher than for red galaxies, which are more difficult to detect at high $z$ due to the shape of their SED. However, since we want to compare red and blue galaxy populations we have to make sure that our sample is not biased towards blue galaxies at high $z$, and we therefore use the conservative stellar mass cut of $\log M_\ast = 9.5$. The colour dependent completeness limits and mass functions of red and blue galaxies in the GNS are further investigated by Mortlock et al. (submitted), who find similar completeness limits. 

In this study we will focus on the redshift range of $1.5 < z < 3$, which provides a co-moving survey volume of $\sim2.3\times10^5$ Mpc$^3$, minimising the effects of cosmic variance. The cosmic variance associated with a certain co-moving volume depends on the clustering strength of the objects and can be roughly estimated from the average galaxy number density and the expected variance of dark matter halos at a certain redshift \citep{Som04}. For the co-moving volume of the GNS within the redshift range of $1.5 < z < 3$, and at the average stellar mass of galaxies in the survey, we expect the influence of cosmic variance to be less than 10\%.
The final galaxy catalogue we use in the following comprises 1289 galaxies down to a stellar mass of $\log M_\ast = 9.5$ within the redshift range of $1.5 < z < 3$.

\section{Galaxy environments}\label{density}

In this study we investigate the environment of each galaxy by computing projected local number densities. We use two different methods to derive local densities: (1) the aperture density, based on galaxy counts in a fixed physical aperture and (2) the nearest neighbour density, based on the distance to the 3$^{rd}$, 5$^{th}$ and 7$^{th}$ nearest neighbour. For both methods we use a redshift interval of $\Delta z = \pm 0.25$ to minimise the contribution of foreground or background objects. As for the stellar masses and rest frame colours, spectroscopic redshifts are used to determine these properties, if available.
The two methods are described in more detail in the following section.

\subsection{Fixed aperture densities}

The fixed aperture densities are measured by counting neighbouring galaxies within a fixed physical radius (or aperture) and redshift interval.
The fixed physical radius was chosen to be 500 $h^{-1}$ kpc, which is roughly the size of a single GNS pointing at $z=1.5-3$.
The number of objects within 500 $h^{-1}$ Mpc co-moving radius at a galaxy's distance and within $\Delta z = \pm 0.25$ is computed for each galaxy, yielding the surface density $\Sigma_{AP}$ in galaxies per Mpc$^2$. 

Edge effects are a major issue due to the design of the GNS, since it does not have a continuous survey area, and edges are encountered frequently. The radius of 500 $h^{-1}$ kpc is not covered for all galaxies, especially for galaxies close to the edges of isolated, non-overlapping pointings, since the field of view of a single pointing covers about 500 kpc at $z > 1.5$. 
To account for the area lost due to survey edges we approximate the surveyed area around each galaxy by the number of image pixels within the aperture, which are actually covered by the observations. $\Sigma_{AP}$ is then normalised by the number of pixels within the aperture. This procedure effectively accounts for differences in aperture area due to galaxies being close to survey edges. 
The normalisation however does not account for variations with redshift due to differences in depth and resolution. To account for those differences we divide each galaxy number density by the median density (denoted by the angled brackets) of all galaxies within a redshift slice of $\Delta z = \pm 0.25$ centred on the galaxy. This then gives the relative overdensity 

\begin{equation}
(1+\rho) = \frac{\Sigma_{AP}}{N_{pixels}} ~ \Big/ \left\langle \frac{\Sigma_{AP}}{N_{pixels}} \right\rangle_{\Delta z}
\end{equation}

\noindent where $\rho$ is the overdensity, i.e. the density excess over the median density. Using the logarithm of $(1+\rho)$ divides the sample in under- and over-dense environments relative to the average density, where by definition $\log~(1+\rho) = 0$.

\subsection{Nearest neighbour densities}

Nearest neighbour densities are widely used in the literature to characterise local galaxy environment. The best choice of the number of neighbours to count, $n$, is open to debate, and depends on the characteristics of the sample and the survey. Values from $n=3$ \citep{Coo06} up to $n=10$ \citep{Dre80} have been used in different studies, depending on the size and mass of the structure that is investigated. A high $n$ is suited for high-mass and high density areas like galaxy clusters, whereas in low density environments a high $n$ measures the distance to other structures rather than characterising the local galaxy density, and therefore a smaller $n$ should be used. However, \citet{Coo05} argue that the choice of $n$ does not change the resulting densities significantly. In this study we measure third, fifth and seventh nearest neighbour densities and compare the results in Section~\ref{results:densities}. 

To account for edge effects a similar technique as used for the aperture densities described above is employed. First, the distance to the third, fifth and seventh nearest galaxy, $D_3$, $D_5$ and $D_7$, within the redshift interval of $\Delta z = \pm0.25$ is computed for each galaxy.
Then, we count the number of image pixels within the area $\pi D_{n}^2$ to approximate the covered area around each galaxy. Since the number of pixels within $\pi D_n$ ($N_{pixels}(D_{n})$) is directly proportional to the covered area, it can be used instead of $\pi D_{n}^2$ to compute the surface density $\Sigma_{n} = n/N_{pixels}(D_{n})$. This gives $\Sigma_{n}$ in arbitrary units, which is ideal for our purposes, since we are only interested in relative densities and $\Sigma_{n}$ is normalised by the median value in each redshift slice, $\langle \Sigma_{n}\rangle_\Delta z$. The nearest neighbour density in units of a relative overdensity is then given by:

\begin{equation}
(1+\delta_n) = \frac{\Sigma_{n}}{ \left\langle \Sigma_{n}\right\rangle_{\Delta z}}
\end{equation}

\noindent where $\delta_n$ itself is the overdensity and $n$ has the values of 3, 5 and 7 respectively. Analogue to the nearest neighbour density we use $\log~(1+\delta_n)$ to distinguish between underdense ($\log~(1+\delta_n) < 0$) and overdense ($\log~(1+\delta_n) > 0$) environments.

\subsection{Monte Carlo simulations}\label{MC}

To estimate the reliability of the local density estimates we perform a set of Monte Carlo simulations. The main source of uncertainty for the local densities are the uncertainties of the photometric redshifts \citep[see e.g.][]{Coo05}, leading to a blurring and loss of information along the line of sight. We therefore randomise the photometric redshift input according to the $\Delta z / (1+z) $ error obtained by the comparison with available secure spectroscopic redshifts in Section~\ref{photoz} and repeat the computation of the local density with the new photometric redshift input.
For this purpose we take the typical photometric redshift error $\Delta z / (1+z) = 0.10$ in our redshift range of $1.5 \leq z \leq 3$ and assume a Gaussian distribution of errors, where the width of the distribution $\sigma$ corresponds to $\Delta z / (1+z) $. For each galaxy a random value is selected within this distribution, which is then added to the measured photo-z.
Galaxies outside our redshift range of interest are included in the randomisation to account for scattering in and out of the redshift range we use. Those galaxies obtain an added value according to the respective $\Delta z / (1+z) $ in their redshift range. We compute an average $\Delta z / (1+z) = 0.08$ for galaxies with redshifts lower than our range of interest ($z < 1.5$) and an average $\Delta z / (1+z) = 0.1$ for all galaxies with redshifts higher than the range we use in this study ($z>3$).
Catastrophic outliers are accounted for by randomly adding much larger offsets to the original redshifts of the percentage of galaxies corresponding to the fraction of catastrophic outliers, which is calculated from the comparison with secure spectroscopic redshifts. Galaxies with $|\Delta z / (1+z)| > 0.3$ are treated as catastrophic outliers here. The offsets are randomly picked from the interval $0.3 < \Delta z / (1+z) < 1$ and added or subtracted from the original redshift.

Following the above procedure we obtain a randomised photometric redshift input catalogue accounting for the average redshift uncertainty within the redshift range of interest as well as for the respective fraction of catastrophic outliers. It also accounts for scattering galaxies in and out of the redshift range we use in this study ($1.5 < z < 3$). The catalogue of randomised photometric redshifts is then used to recompute the local densities, aperture and nearest neighbour densities, as described above. The whole procedure is repeated 100 times, obtaining 100 randomised local density estimates, from which we then compute the average local density uncertainty. The local density error for each individual galaxy is given by the standard deviation of all Monte Carlo runs. These individual errors are then averaged to obtain an average uncertainty of the respective local density estimator.
We obtain the following average uncertainties for the different local density estimates: $\Delta \log~(1+\delta_3) = 0.24$, $\Delta \log~(1+\delta_5) = 0.21$, $\Delta \log~(1+\delta_7) = 0.18$ for the three nearest neighbour densities and $\Delta \log~(1+\rho) = 0.20$ for the aperture density. These values of around 0.2 dex show the we can measure local densities with roughly the same or even slightly better accuracy than stellar masses (see Section~\ref{stellar mass and colours}).
The results of the Monte Carlo simulations are also discussed in the context of the relations between colour density and stellar mass in the following sections, and are shown in the respective figures.

\begin{figure*}
\includegraphics[width=0.485\textwidth]{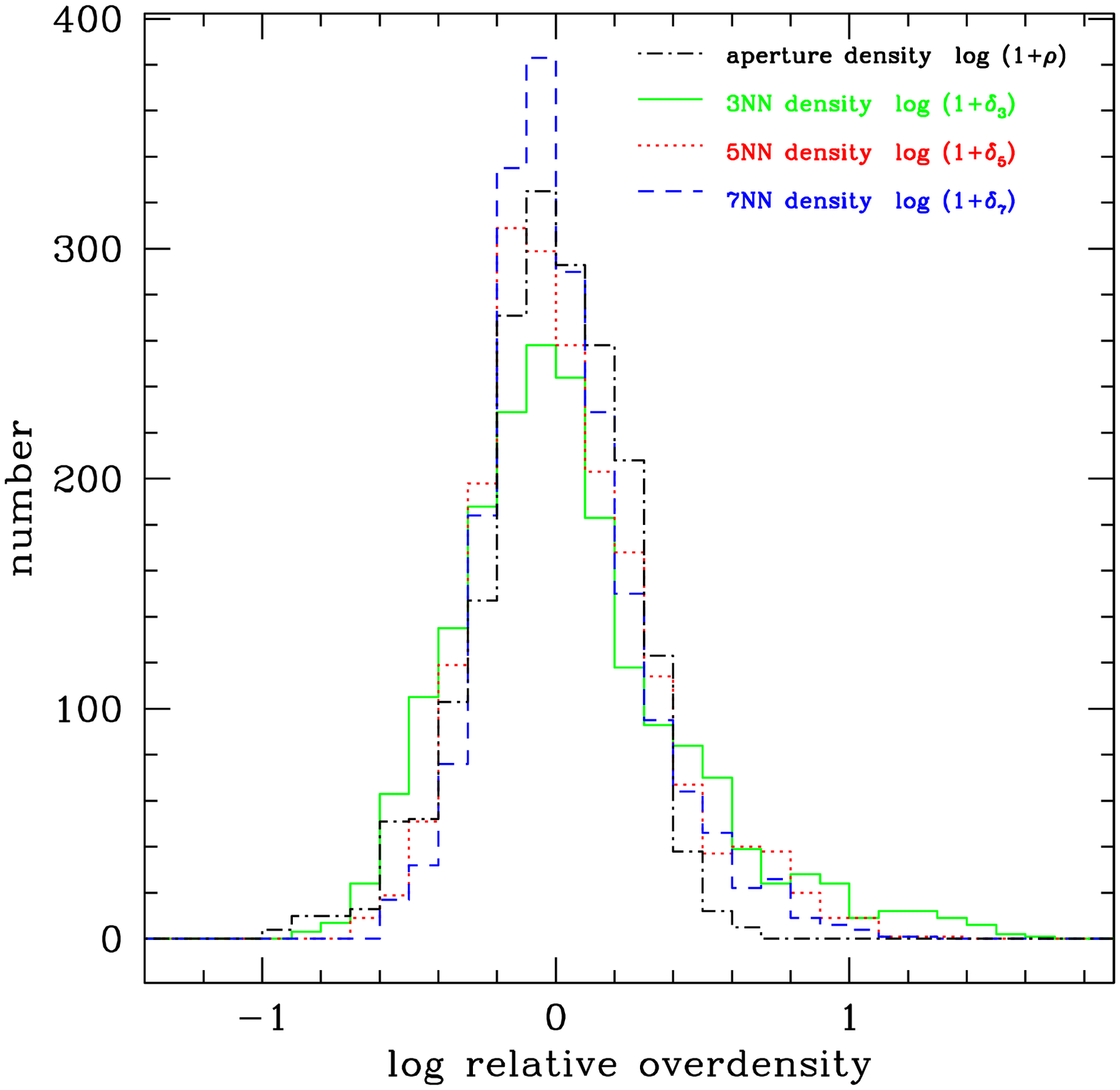}
\includegraphics[width=0.485\textwidth]{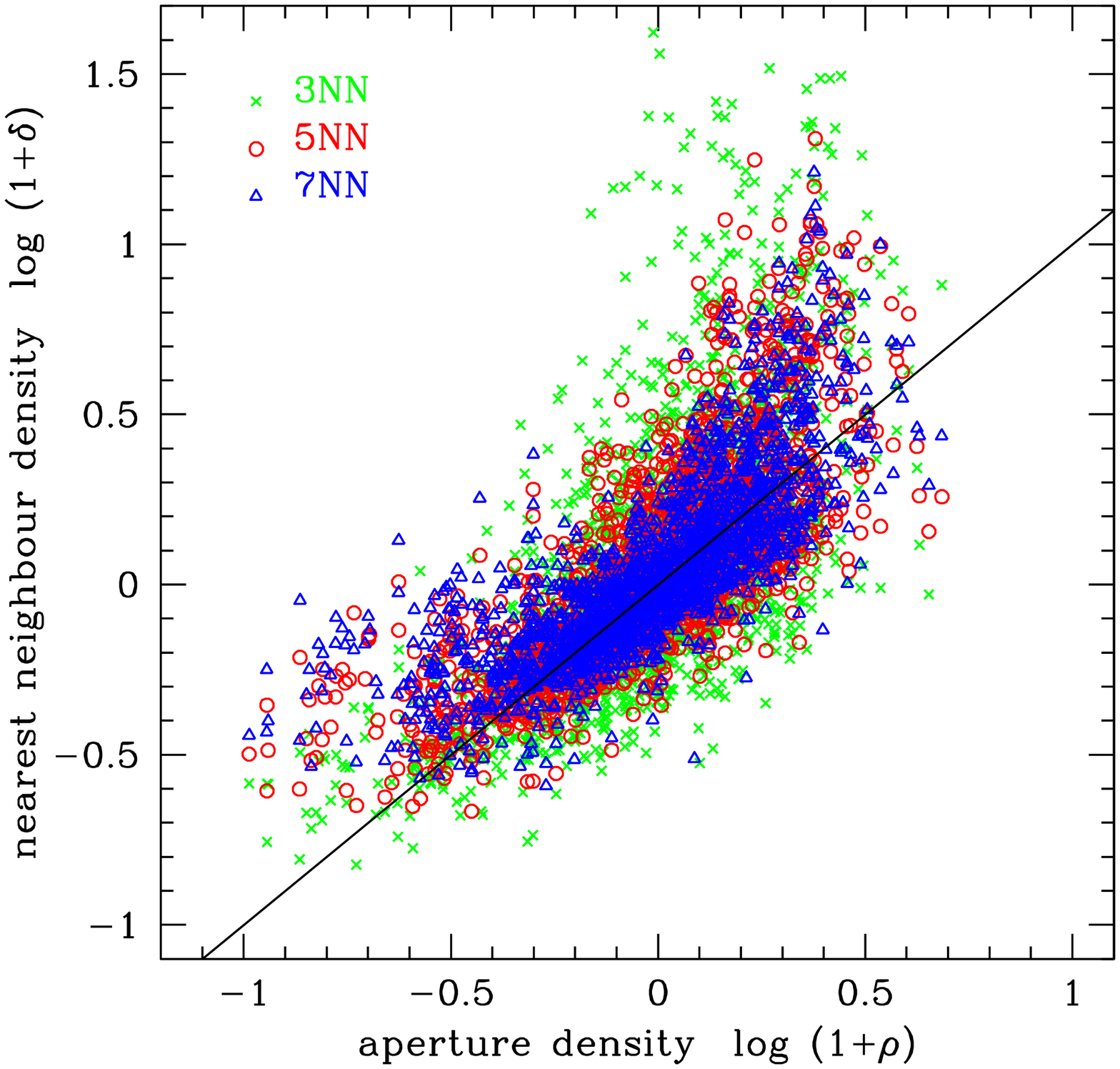}
\caption{Comparison between local density estimators. Left panel: Histogram of nearest neighbour densities $\log~(1+\delta_n)$ in green ($n=3$), red dotted ($n=5$) and blue dashed ($n=7$) and fixed aperture density $\log~(1+\rho)$ in black, dashed-dotted line. Right panel: $\log~(1+\delta_n)$ versus $\log~(1+\rho)$. Nearest neighbour densities are plotted as green crosses ($n=3$), red circles ($n=5$) and blue triangles ($n=7$) respectively. The solid line is the one-to-one relation.
\label{fig4}}
\end{figure*}

\begin{figure*}
\includegraphics[width=0.325\textwidth]{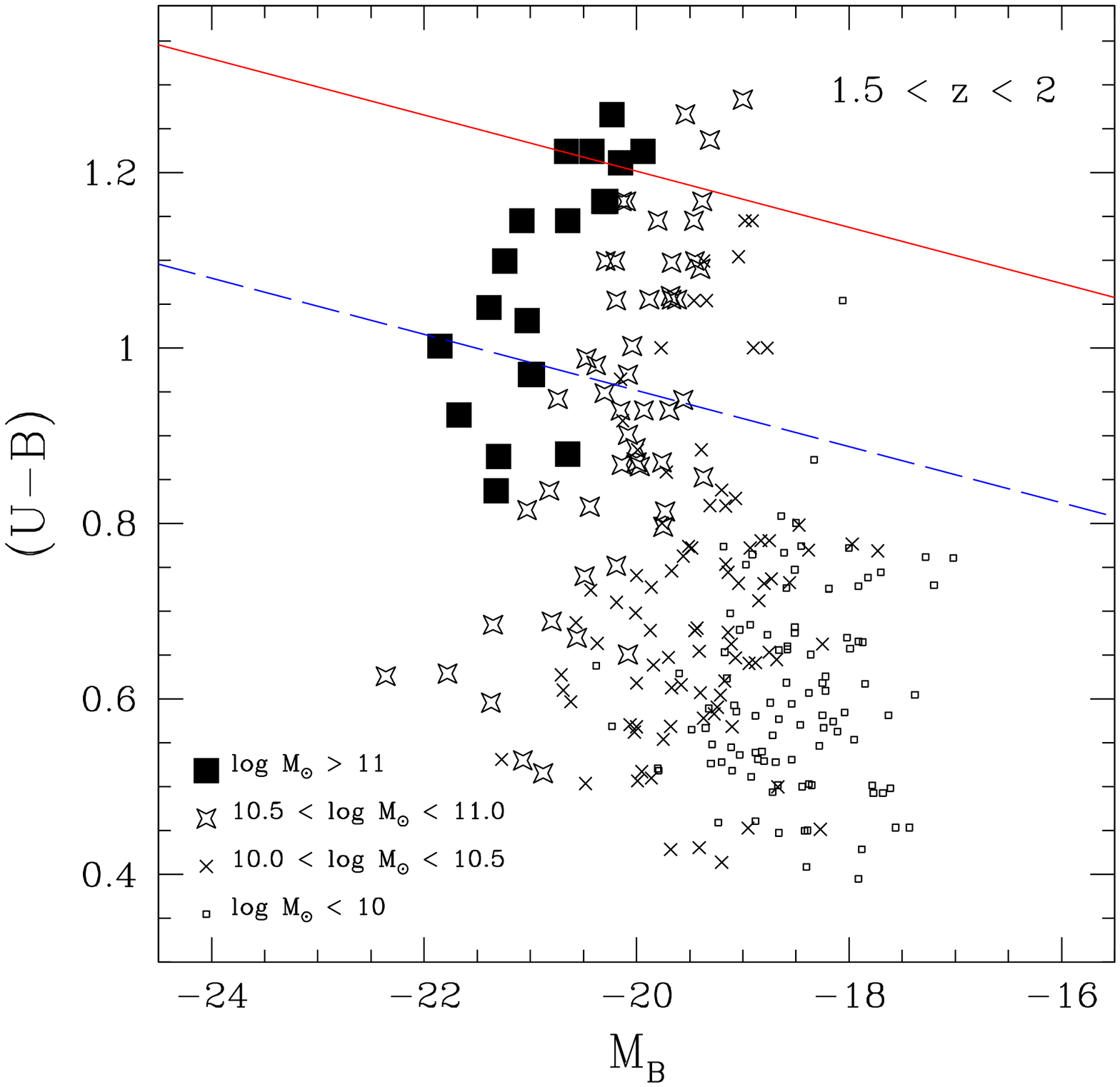}
\includegraphics[width=0.325\textwidth]{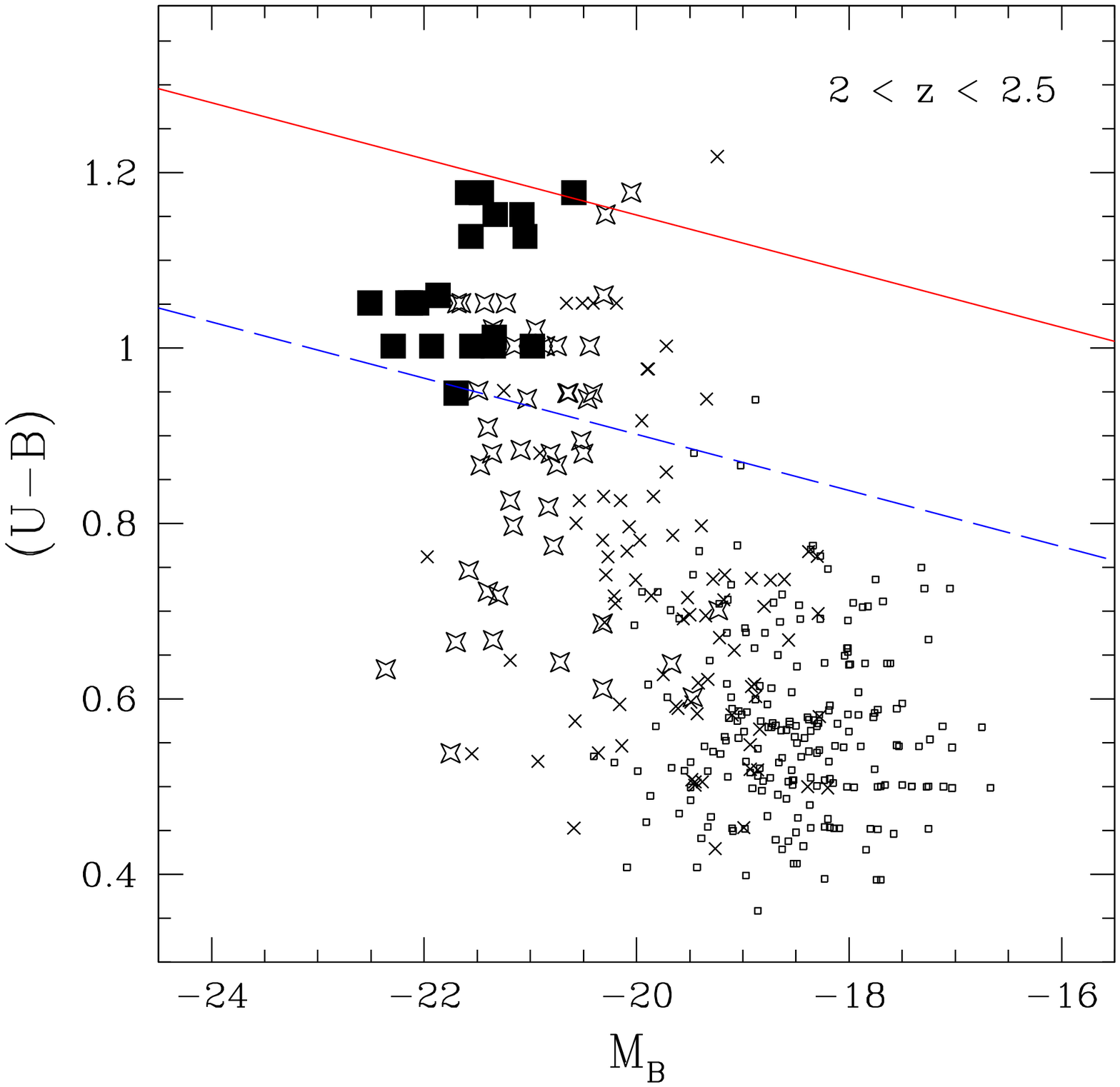}
\includegraphics[width=0.325\textwidth]{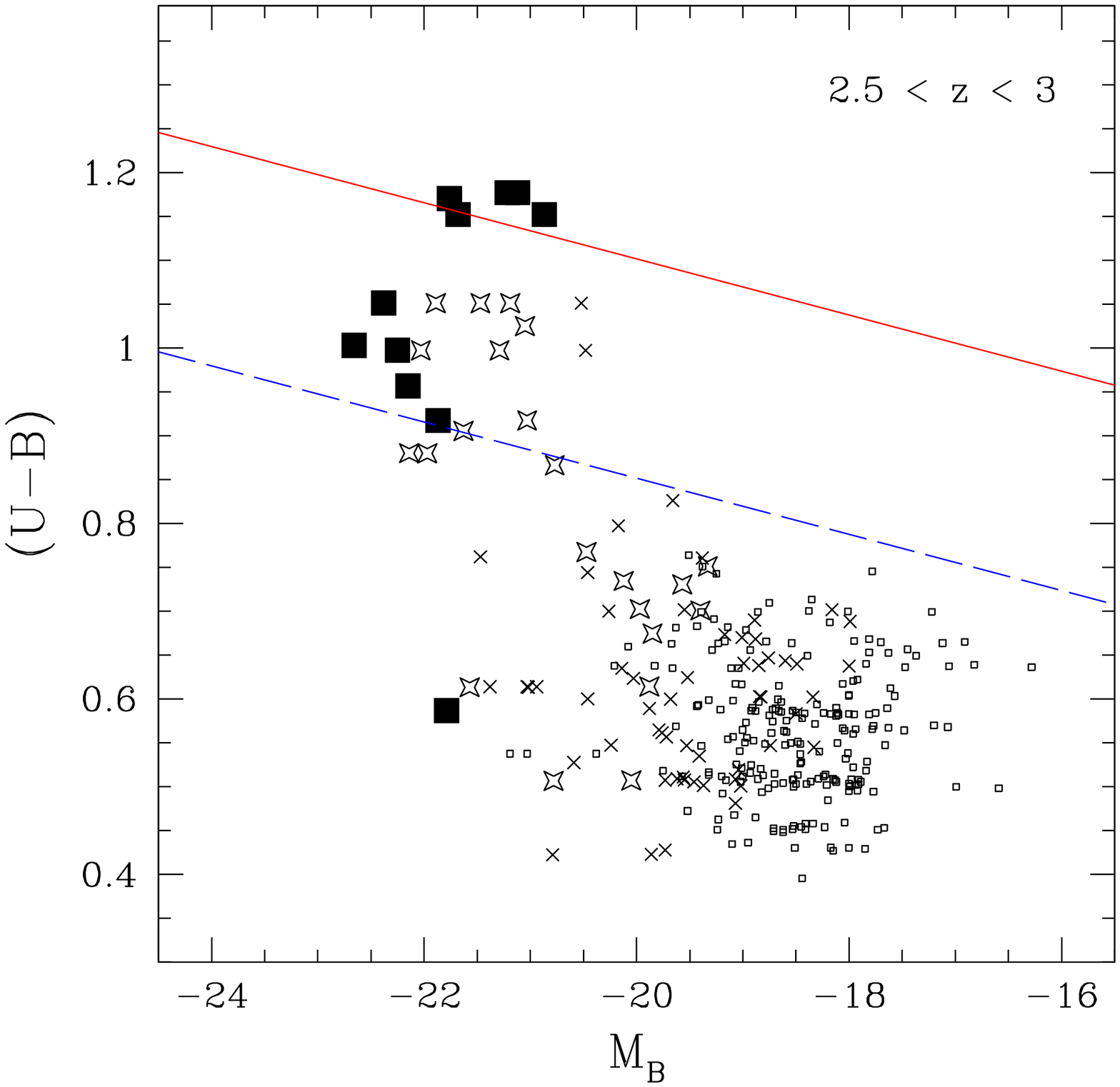}
\includegraphics[width=0.325\textwidth]{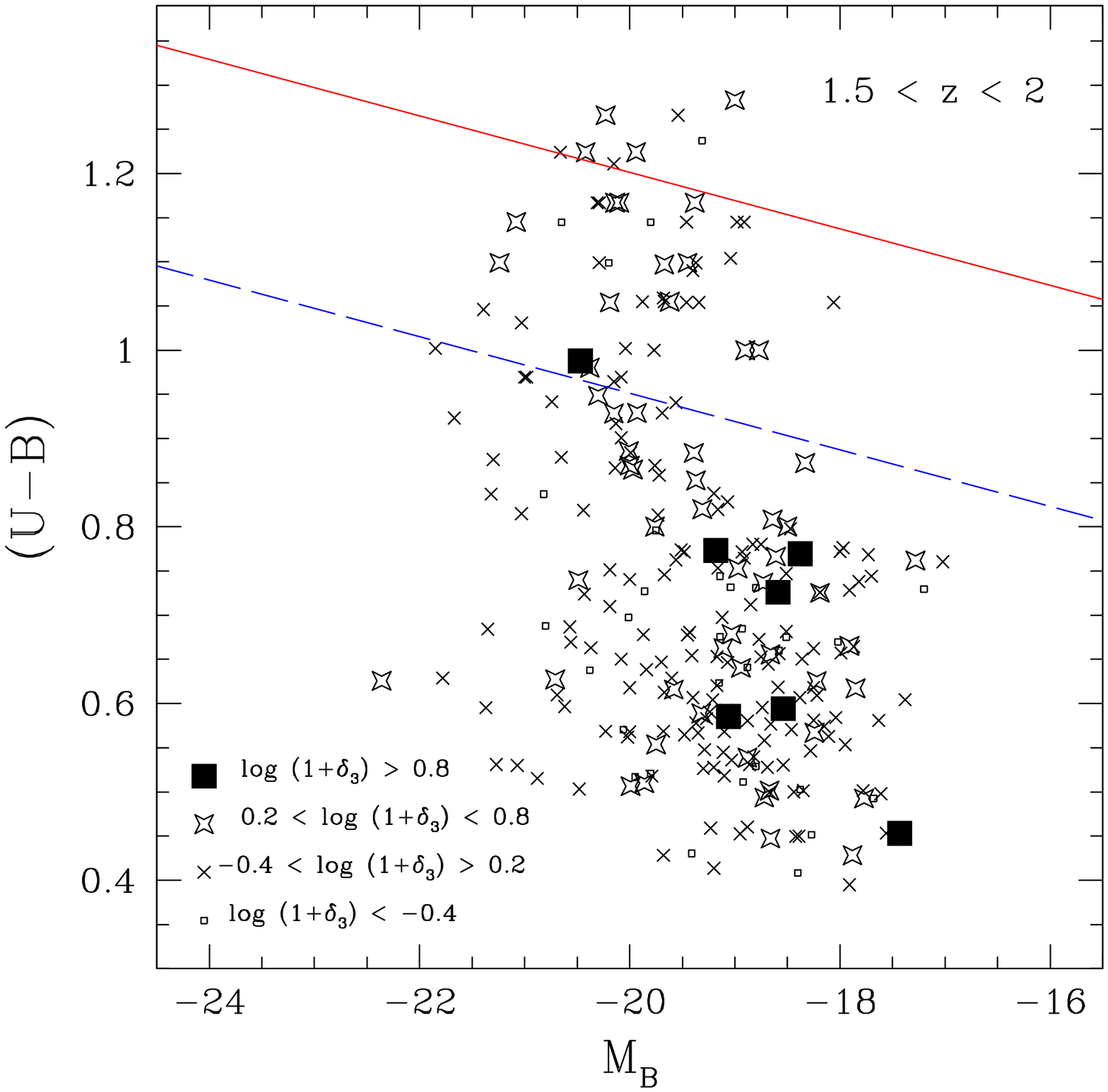}
\includegraphics[width=0.325\textwidth]{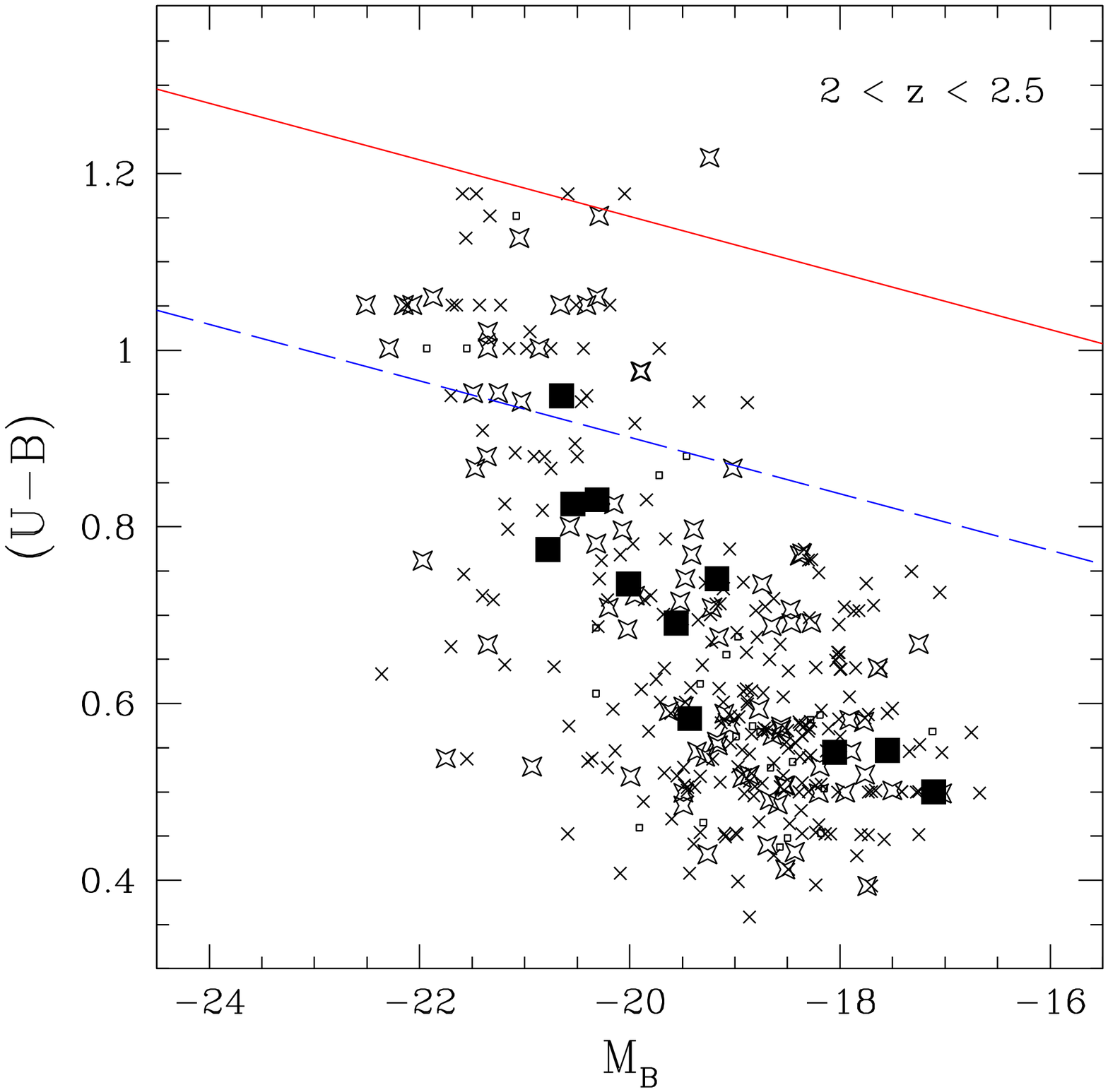}
\includegraphics[width=0.325\textwidth]{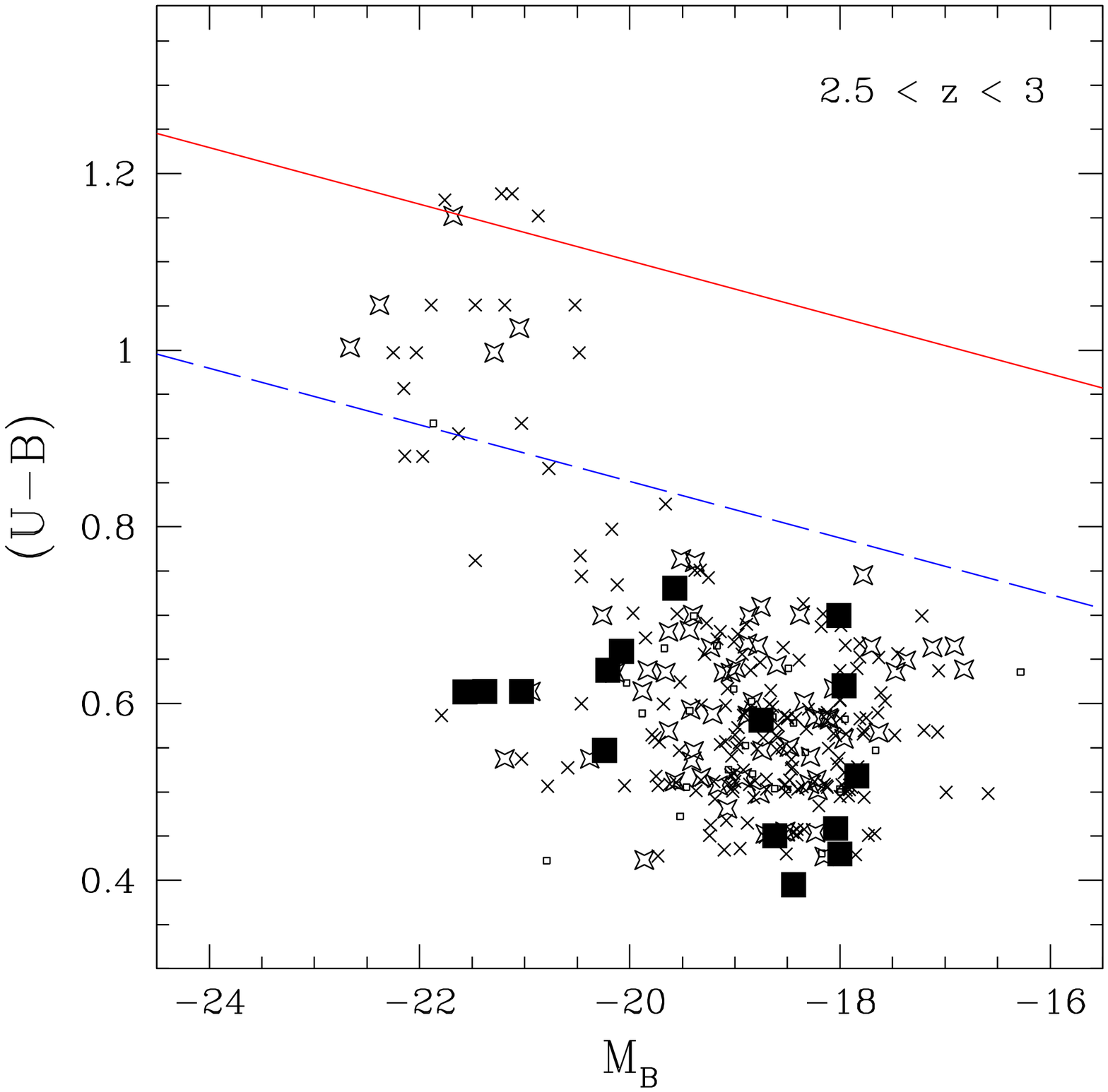}
\caption{Colour-magnitude relation in three redshift ranges, divided in bins of stellar mass (top panels) and local density (bottom panels). The symbols are shaped and sized according to the stellar mass bin (top) and density bin (bottom) in which the galaxy lies. The larger symbols correspond to higher stellar mass in the top panel, and to higher overdensity in the bottom panel. Top panels: CMR in different stellar mass bins: $9.5 < \log M_\odot < 10$ (open circles), $10 < \log M_\odot < 10.5$ (crosses), $10.5 < \log M_\odot < 11$ (stars), and $\log M_\odot > 11$ (solid squares). Bottom panels: CMR in different local density bins: $ \log~(1+\delta_3) < -0.4 $ (open circles), $-0.4 < \log~(1+\delta_3) < 0.2$ (crosses), $0.2 < \log~(1+\delta_3) < 0.8$ (stars), and $\log~(1+\delta_3) > 0.8$ (solid squares). The expected location of the red sequence at the respective redshift is shown as red solid line, while the limit to the blue cloud below which a galaxy is considered blue is plotted as blue long-dashed line. \label{fig5}}
\end{figure*}

\section{Results} 

In the following we first discuss the correlations between the local galaxy densities $\log~(1+\delta_n)$ and $\log~(1+\rho)$. We then investigate the colour-magnitude relation in bins of stellar mass and bins of local density, and show the behaviour of rest-frame colour $(U-B)$ and blue fraction of galaxies as a function of both local density ($\log~(1+\delta_3)$) and stellar mass ($\log~M_\ast$).
Additionally, we test differences between galaxy properties in the high and low quartiles of the local density and stellar mass distributions respectively.
We focus on the redshift range of $1.5 \leq z \leq 3$, which has a reasonable surveyed volume and stellar mass completeness down to $\log~M_\ast = 9.5$.

\subsection{Galaxy local densities}\label{results:densities}

As described above two different methods were used to estimate local galaxy densities, the fixed aperture density, $(1+\rho)$, and the $3^{rd}$, $5^{th}$ and $7^{th}$ nearest neighbour density, $(1+\delta_3)$, $(1+\delta_5)$ and $(1+\delta_7)$. Figure~\ref{fig4}, left panel, shows the distribution of the different measures of relative overdensities: $\log~(1+\rho)$ is plotted as a black dashed-dotted line and $\log~(1+\delta_n)$ as a green solid line ($n=3$), red dotted line ($n=5$) and blue dashed line ($n=7$). 
The figure shows the different dynamical range of the four methods: the nearest neighbour densities $\log~(1+\delta_n)$ cover a wider range of relative densities, especially at the high density end. The $3^{rd}$ nearest neighbour density has the widest range, since it is most sensitive to density variations on very small scales. The standard deviation of the $\log~(1+\rho)$ distribution is $\Delta \log~(1+\rho)=\pm0.26$, while $\Delta \log~(1+\delta_3) = \pm0.39$.  The good overall agreement between the four methods is shown in the right panel of Figure~\ref{fig4}, where $\log~(1+\delta_n)$ is plotted against $\log~(1+\rho)$. The different symbols used are green crosses ($n=3$), red circles ($n=5$) and blue triangles ($n=7$), while the one-to-one relation is shown as a black solid line. 
The four density estimates agree well on average, with an average spread of $\sim0.2-0.3$ dex. This scatter reflects the average uncertainty of the local density measurements of $\sim0.2$ dex obtained by the Monte Carlo simulations described in Section~\ref{MC}. However, it also shows that the different methods trace galaxy environment on different scales and are not necessarily comparable. 
The different sensitivity of the different values of $n$ is visible at the low density end, where the higher $n$ traces the average densities on larger scales, rather than local underdensities, which we are interested in. At densities above $\log~(1+\rho) \sim -0.5$ the four methods agree best on average, while the higher sensitivity of densities measured with lower $n$ towards galaxy concentration on small scales becomes visible: high relative overdensities measured with $n=3$ (green crosses in Figure~\ref{fig4}) are partly recovered by $n=5$ but not detected using $n=7$ or the fixed aperture density.
The $n^{th}$ nearest neighbour density is by construction a better tracer of the extremes in the local density distribution, since it uses an adaptive area and therefore is sensitive to galaxy concentrations at very small scales, that the fixed aperture cannot probe. The lower the value of $n$, the more sensitive the density measure is to local galaxy concentrations on small scales. These very local galaxy concentrations however are potentially influencing the evolution of a galaxy dramatically via merging or galaxy interactions.

In the following we use the $3^{rd}$ nearest neighbour density $\log~(1+\delta_3)$ as a local density estimator, since - although it has slightly larger uncertainties than the other density estimates (see Section~\ref{MC}) - it has a wider dynamical range and higher sensitivity to the most underdense and overdense environments and is therefore a better probe of the local environment.

\subsection{The colour-magnitude relation at different stellar masses and local densities}\label{colour-magnitude}

In the following we investigate the colour-magnitude relation of the GNS sample in different redshift ranges. We investigate if the red sequence is already present at $z>1.5$ and if it is populated by high or low-mass galaxies. Additionally we examine if the build-up of the red sequence preferentially takes place in a certain environment. For this purpose the sample is split up in four stellar mass bins and four local density bins.

Figure~\ref{fig5} shows the $(U-B)$ - $M_B$ colour-magnitude relation (CMR) in three different redshift bins: $1.5<z<2$, $2<z<2.5$ and $2.5<z<3$. In the three top panels the sample is split in 4 stellar mass bins with a bin size of 0.5 dex; the symbols are shaped and sized according to the galaxy's stellar mass: larger symbols represent higher stellar mass galaxies. In the three lower panels the sample is split in bins of local density with a bin size of 0.6 dex. Here, larger symbols correspond to galaxies located in higher overdensities.
The red solid lines indicate the expected location of the red sequence observed at $z\sim1$, and evolved passively back in time according to the models of \citet{vDF01}, as described in Section~\ref{completeness-colour}. The blue long-dashed lines is the border to the blue cloud, defined as being 0.25 magnitudes bluer than the red sequence (see Section~\ref{completeness-colour}). 

The different slices in stellar mass are clearly separated in colour-magnitude space. The separation is perpendicular to the red sequence, and the distance from the expected red sequence increases with decreasing stellar mass. 
The different slices in local density on the other hand are not well separated in colour-magnitude space. The relation between colour and local density is much weaker than the relation between colour and stellar mass. This subject is further discussed in Section~\ref{colour-mass-density}.

 \begin{figure}
\includegraphics[width=0.48\textwidth]{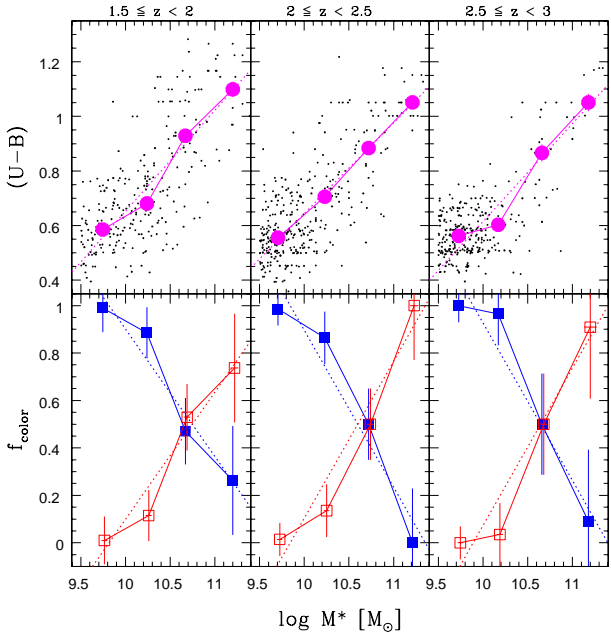}
\caption{Colour-stellar mass relation in three redshift bins. The mean and error of the mean in each bin of stellar mass (the bin size is 0.5 dex) is overplotted in magenta (big symbols). Blue fractions are plotted as blue solid squares, red fractions as red open squares.  The dotted lines show least squares fits to the data. 
\label{fig6}}
\end{figure}

As can be seen, massive red galaxies are already present at a redshift of $2.5 < z < 3$, when the universe was only $\sim$2 Gyrs old. The location of these galaxies in the colour-magnitude diagram is consistent with the location expected from passively evolving back the red sequence found at $z\sim1$ (see above). At $z>2.5$ this region in colour-magnitude space is populated mainly by the most massive galaxies ($\log~M_\ast > 11$), while at lower $z$ the range of galaxies on or close to the red sequence extends towards lower mass galaxies. Note that these most massive galaxies are mostly star-forming (Bauer et al., submitted) and not ``red and dead" as their location in the colour-magnitude diagram close to the red sequence might suggest.

Galaxies close to or on the location of the red sequence span a wide range of local densities, and there is no clear preference for a certain environment, however, we do not see red galaxies uniquely located in the most overdense regions. This is opposite to what was found at lower redshifts up to $z=1$ \citep[see e.g.][]{Gru10} and might hint at an inversion of the colour-density relation at high $z$, as discussed later in this paper (Section~\ref{colour-mass-density} and Section~\ref{highlow}). 

Clear evolution in $M_B$ is seen for the galaxies on or close to the expected red sequence, while the blue cloud remains basically unchanged. Part of this evolution, as well as the lack of low-mass galaxies at high $z$ may be due to the fact that faint red galaxies become more difficult to detect with high $z$ than faint blue galaxies and we might miss some low surface brightness, low-mass red galaxies in our sample. However, the mass limit we adopt in this study is very conservative and our sample should not be affected by colour-dependent incompleteness. A more detailed discussion on the stellar mass completeness and stellar mass functions for blue and red galaxies will be presented in Mortlock et al. (submitted).

\subsection{Colour-mass and colour-density relations}\label{colour-mass-density}

A strong relation between rest-frame colour and stellar mass and the apparent lack of a strong relation between colour and local density was found in the colour-magnitude relation in the previous section. We now investigate these correlations in more detail. The same redshift ranges as already used for the analysis of the colour-magnitude relation are used here: $1.5<z<2$, $2<z<2.5$ and $2.5<z<3$. To quantify the correlation between two variables we compute Spearman rank correlation coefficients $R$. The value of $R$ ranges between $-1 \leq R \leq 1$, where $R=1$ ($R=-1$) means that two variables are perfectly correlated (anti-correlated) by a monotonic function. Completely uncorrelated variables result in $R=0$. Taking into account the sample size in each bin we estimate the significance of the value of $R$ using the conversion from correlation coefficient to z-score \citep{Fie57}.

Figure~\ref{fig6}, shows rest-frame $(U-B)$ colour (top) and the fractions of blue and red galaxies (bottom) as a function of stellar mass in three redshift bins. The correlation between $(U-B)$ colour and stellar mass $\log~M_\ast$ is highly significant at all redshifts. Spearman rank correlation coefficients in the three redshift bins range between 0.7-0.8, corresponding to a significance of $\sim 6\sigma$ at all redshifts.
The colour-mass relation also does not appear to evolve strongly with redshift. The cross-over mass, i.e. the stellar mass at which there is an equal number of red and blue galaxies, stays roughly constant at $\log~M_\ast \sim 10.8$. This is the same cross-over mass found at lower redshifts $0.4<z<1$ \citep{Gru10}. Note that in the computation of blue and red galaxy fractions we account for the passive reddening of galaxy colours with redshift (see Section~\ref{completeness-colour}). 

The fraction of blue galaxies over all stellar masses evolves with redshift from $f_{blue} \sim 0.95$ at $z\sim3$ to $f_{blue} \sim 0.55$ at $z<1$. Notice that the fading and reddening expected from passive evolution of a galaxy's stellar population is already accounted for in the definition of red and blue galaxies at each redshift. Combining this with the above considerations of almost no evolution in the colour-stellar mass relation, the colour evolution is mainly taking place at low stellar masses of $\log~M_\ast < 10.5$. The population of more massive galaxies already comprises a similar fraction of red galaxies at high redshifts up to $z\sim3$ compared to $z<1$. 

\begin{figure*}
\includegraphics[width=0.485\textwidth]{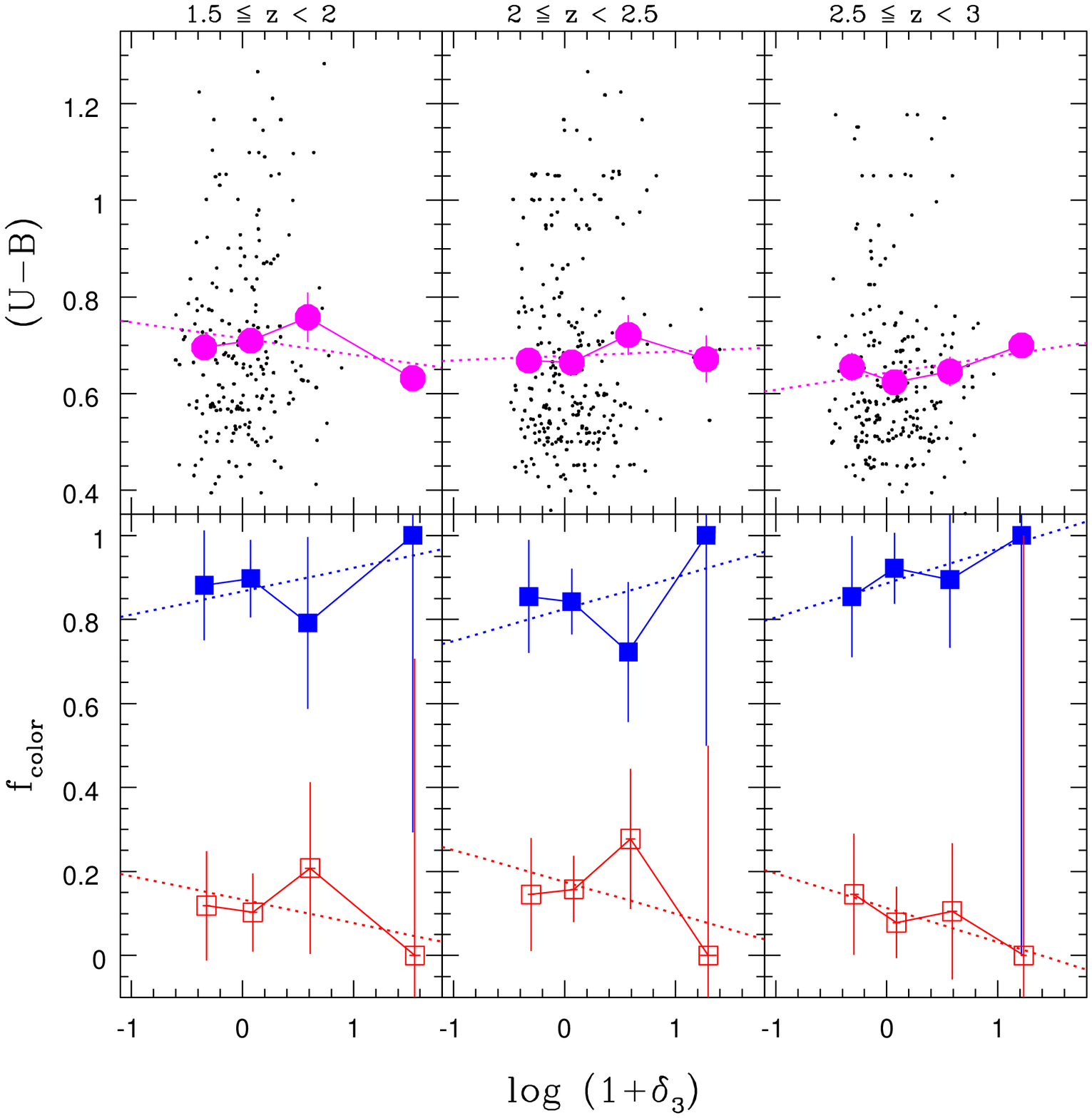}
\includegraphics[width=0.485\textwidth]{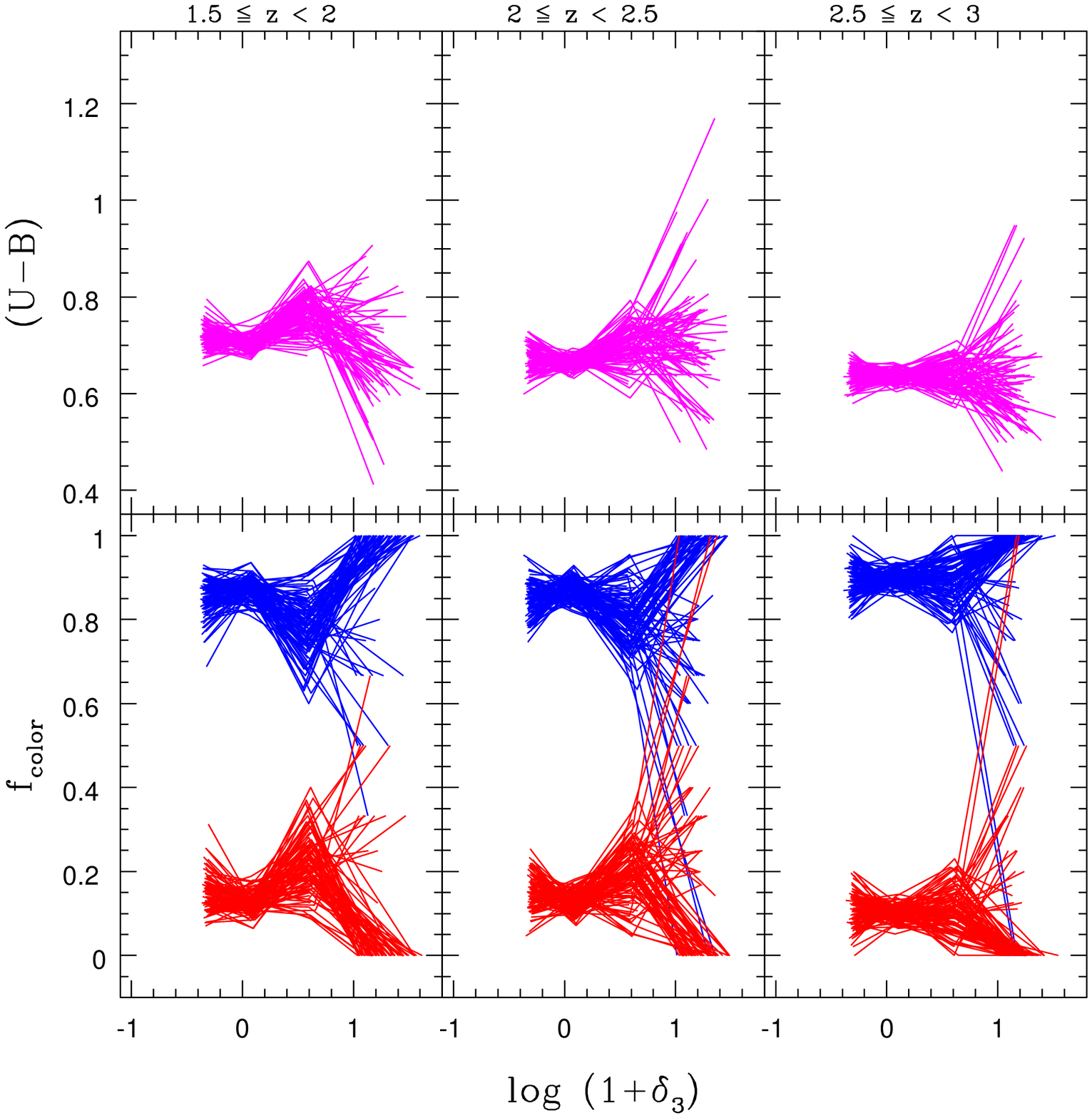}
\caption{Colour-density relation in three redshift ranges for the whole mass range ($\log~M_\ast > 9.5$). The redshift bins are indicated on top of each panel.
Left panel: $(U-B)$ colour and fraction of red and blue galaxies as a function of local density $\log~(1+\delta_3)$. The mean and error of the mean in each bin of local density (the bin size is 0.6 dex) is overplotted in magenta (big symbols). Blue fractions are plotted as blue solid squares, red fractions as red open squares.  The dotted lines show least squares fits to the data. 
Right panel: Monte Carlo simulation of the data: local densities recomputed 100 times based on randomised photometric redshift input (see text). The colour coding is the same as in the left hand panel.
\label{fig7}}
\end{figure*}

The left panel of Figure~\ref{fig7} shows the relation between $(U-B)$ colour (top row) and the fraction of blue and red galaxies (bottom row) as a function of relative overdensity $\log~(1+\delta_3)$. We do not find a significant correlation between colour and overdensity in the data. There is, however, a possible trend for a higher fraction of blue galaxies ($\sim 100\%$) at the highest overdensities ($\log~(1+\delta_3) > 0.8$) at all redshifts, where as the blue fraction at intermediate and low densities is around 80-90\%. 
The trend of a lack of red galaxies at high overdensities noticed in the CMR in Figure~\ref{fig5} is visible, but not significant due to the low number of galaxies at high overdensities.

To further check the significance of the higher fraction of blue galaxies at high overdensities we use Monte Carlo simulations. We perform 100 Monte Carlo runs based on the randomised photometric redshift input as described in Section~\ref{MC}. The scatter between the individual runs represents the uncertainties in the local densities due to the photometric redshift error. The right panel of Figure~\ref{fig7} shows the result of the simulations. The average colour (top panels) and fraction of blue and red galaxies (bottom panels) as a function of local density in each Monte Carlo run is plotted. The large scatter in the high density end due to the low number of galaxies in extreme overdensities becomes visible. 

Although, above densities of $\log~(1+\delta_3) > 0.8$, and at all redshifts, the vast majority of Monte Carlo runs show a blue fraction of 100\%, whereas at lower densities the blue fractions vary between 85-90\%. The average and RMS of all Monte Carlo runs at $\log~(1+\delta_3) < 0.8$ shows an average blue fraction of 85$\pm$2\% in the two lower $z$ bins and 90$\pm$2\% in the high $z$ bin. At the highest densities the average blue fraction is 95$\pm$13\%, however, the distribution of average blue fractions at the highest overdensities is not Gaussian and the standard deviation does not represent the distribution adequately. 
To estimate the probability that the difference between high and low density is caused by chance, we consider the number of times a Monte Carlo run gives a blue fraction at the highest densities which is lower than the average value plus 3$\sigma$ at lower densities. In the lowest redshift bin about 90\% of Monte Carlo runs result in $f_{blue} >90\%$, giving a $\sim10\%$ probability that the difference between highest and lower densities is caused by chance. The intermediate and high redshift bin have a probability of $\sim15\%$.

We conclude that the colour-density relation at $z>1.5$ is practically not existent or very weak. If a trend with local density persists it must be very minor, with a variation of the average $(U-B)$ colours of less than $\sim$0.1 magnitudes between relative local densities that differ by a factor of 100. There could be an environmental influence on the blue fractions of galaxies in the most extreme overdense environments. This difference in blue fractions amounts to about 10\% and is not detectable at a statistically significant level with the low number of galaxies in our sample located at the highest overdensities (33 galaxies at $\log~(1+\delta_3) > 0.8$ between $1.5 < z < 3$).

\begin{figure*}
\includegraphics[width=0.75\textwidth]{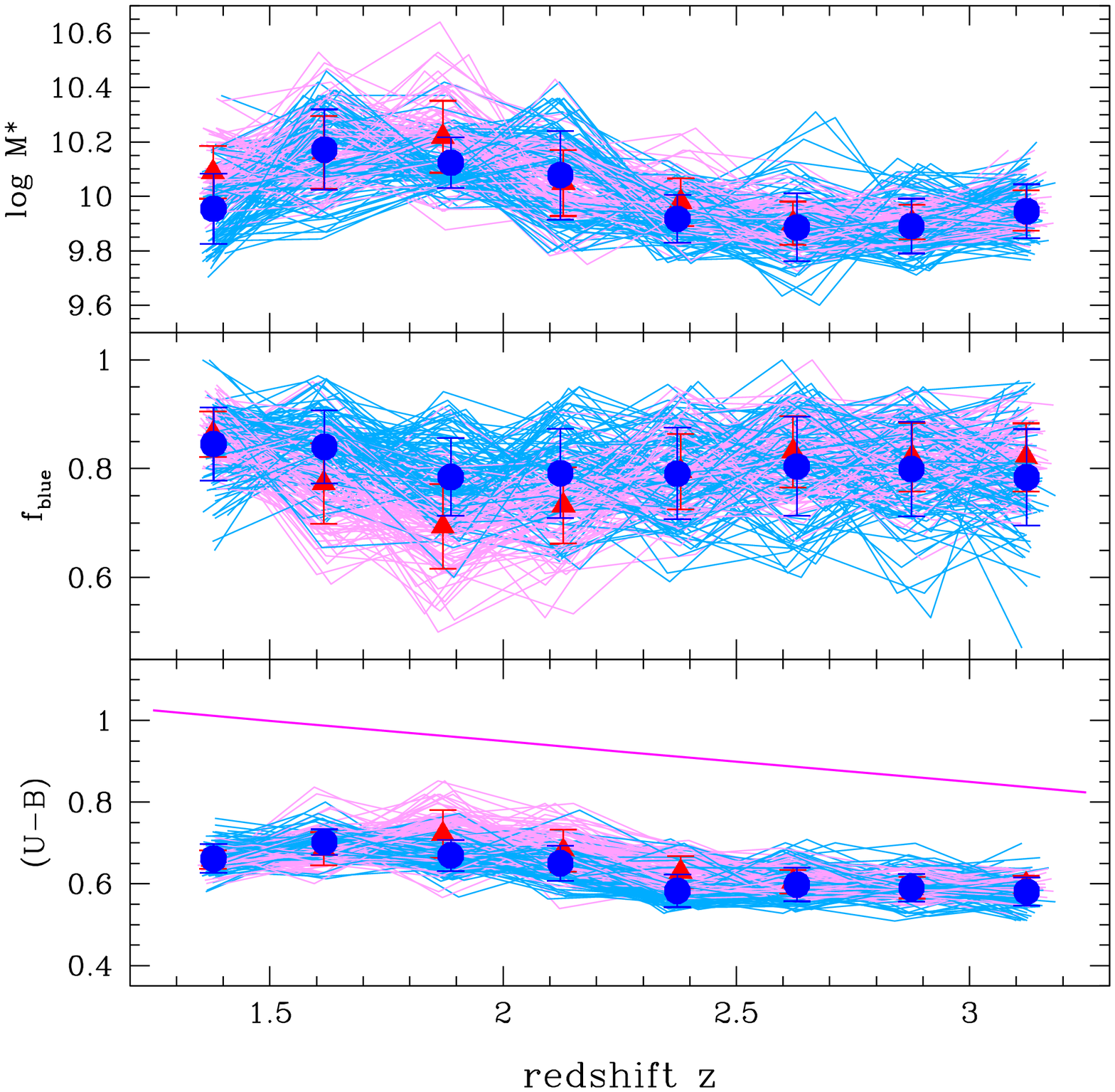}
\caption{Colour, blue fraction, and stellar mass in the high and low quartiles of the local density distribution ($\log~(1+\delta_3)$). The high quartile corresponds to $\log~(1+\delta_3)>0.22$, the low quartile to $\log~(1+\delta_3)<-0.23$.  Each line is the result of one Monte Carlo run. The average and RMS of all Monte Carlo runs in each redshift slice of 0.25 is overplotted as big symbols. Red lines and triangles correspond to the high density quartile, while blue lines and circles correspond to the low quartile of the density distribution. The magenta straight line in the bottom panel indicates the distinction between red  and blue galaxies at $M_B = -21.5$, calculated from evolving back the red sequence as described in Section~\ref{completeness-colour}.
\label{fig8}}
\end{figure*}

\begin{figure*}
\includegraphics[width=0.75\textwidth]{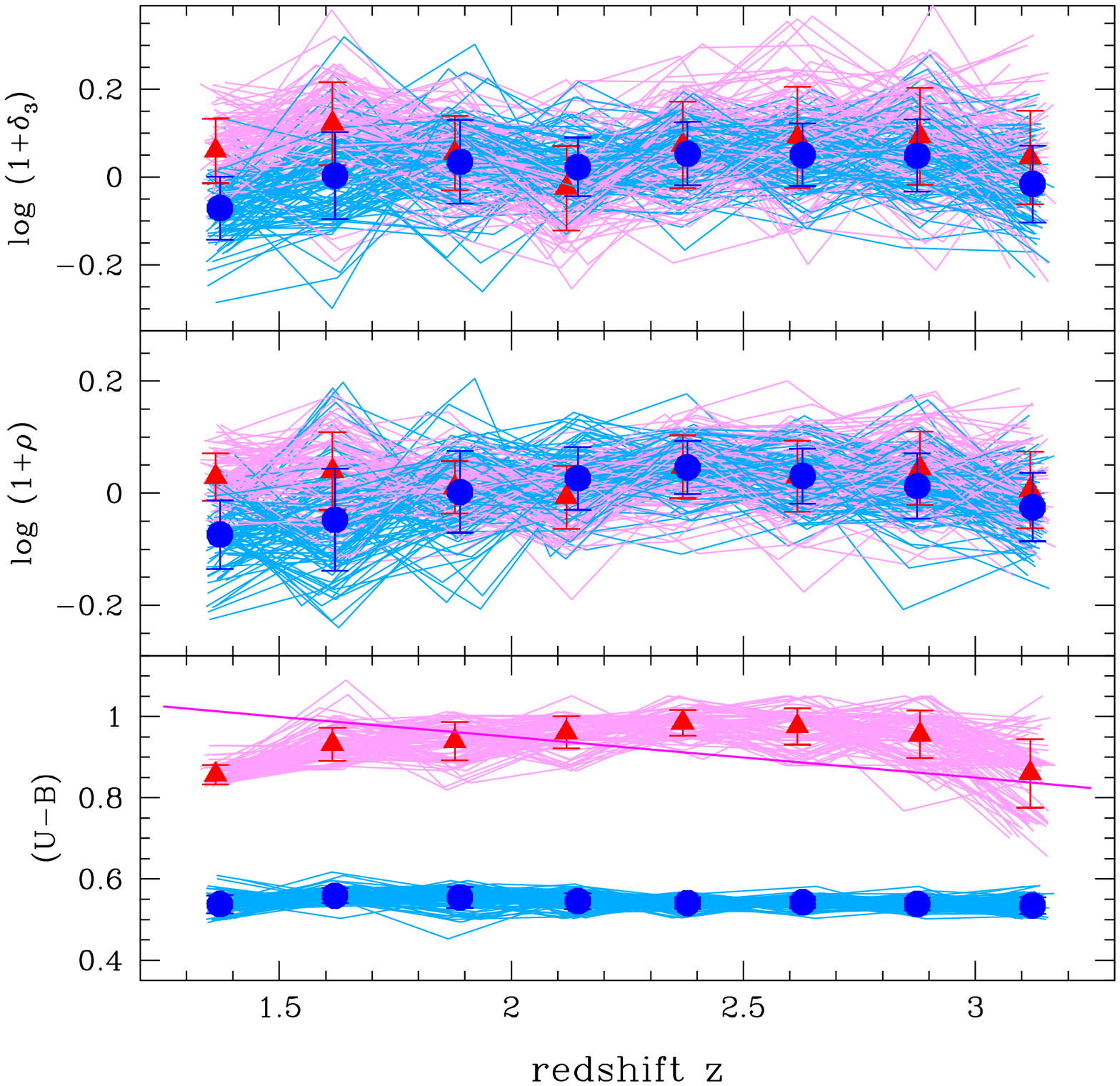}
\caption{Local density ($\log~(1+\delta_3)$ and $\log~(1+\rho)$) and $(U-B)$ in the high and low quartiles of the stellar mass distribution. The high quartile corresponds to $\log~M_\ast > 10.5$ (red), and the low quartile to $\log~M_\ast< 9.7$ (blue). Symbols and lines are the same as in Figure~\ref{fig8}.\label{fig9}}
\end{figure*}

\subsection{Galaxy properties in the high and low quartiles of density and mass distribution}\label{extreme}

In the following we compare the properties of galaxies in the low and high quartiles in the distribution of local density and stellar mass, i.e. 25\% of galaxies at each extreme respectively. The high quartile of local density corresponds to galaxies located in overdense regions of $\log~(1+\delta_3) > 0.22$, while the low quartile corresponds to underdense regions of $\log~(1+\delta_3) < -0.23$. The stellar mass distribution has its low quartile at $\log~M_\ast < 9.7$ and the high quartile at $\log~M_\ast > 10.5$.

After dividing the sample into low and high quartiles a Kolmogorov-Smirnov (K-S) test between the different quartiles is used to investigate the statistical difference of the two subsamples. To increase the number statistics we consider the full redshift range of our sample at $1.5 \leq z \leq 3$, instead of binning into the three redshift ranges used above. The stellar mass limit is kept at $\log~M_\ast = 9.5$, as above. 
The K-S test gives the probability $P_{low,high}$ that the galaxy properties in the low and high quartile come from the same parent distribution. The statistical differences we find over the whole redshift range ($1.5 \leq z \leq 3$) can be summarised as: \\
(1) galaxies in the low and high quartile of the local density distribution have marginally different colours ($P_{low,high}=0.05$, i.e. $\sim 2 \sigma$) but not statistically different stellar masses ($P_{low,high}=0.12$, i.e. $< 2 \sigma$); and \\
(2) galaxies in the low and high quartile of the stellar mass distribution have statistically different colours ($P_{low,high}<0.0001$, i.e. $> 3 \sigma$), as well as marginally different local densities ($P_{low,high}=0.02$, i.e. $> 2 \sigma$).

Although the differences between the low and high stellar mass quartiles are much clearer than the differences in the local density quartiles, there is a difference at the 2 $\sigma$ confidence level between the colours of galaxies in over and under-dense environments. We further investigate the reliability of this result by using the Monte Carlo simulations described in Section~\ref{MC}. 

In Figure~\ref{fig8} and Figure~\ref{fig9} we plot galaxy properties in the low and high quartiles of local density and stellar mass distribution, respectively, as a function of redshift. The results of each of the 100 Monte Carlo runs are plotted as a single line. 
Figure~\ref{fig8} shows the median $(U-B)$ colour, blue fraction and median stellar mass in each redshift bin ($\Delta z = 0.25$) in the low quartile (blue) and high quartile (red) of local densities. Figure~\ref{fig9} shows median $(U-B)$ colour, and two different densities (fixed aperture density $\log~(1+\rho)$ and third nearest neighbour density $\log~(1+\delta_3)$) in the low (blue) and high quartile (red) of stellar mass. 
The average of all Monte Carlo runs in each redshift slice is overplotted as big symbols, blue circles for the low quartile and red triangles for the high quartile.

Figure~\ref{fig8} shows that there is barely any difference in colour, blue fraction or stellar mass between the low and high quartile of local density. We may see evidence for a weak trend of a lower blue fraction at high densities emerging at $z < 2.2$, as well as a correspondent peak in the average stellar mass in the high density quartile, which is probably responsible for the positive correlation found from the K-S test on the data described above. However, we conclude that we lack the number statistics to reliably confirm the presence of a difference between low and high density quartile and that our data is consistent with there being no difference.

Note however that the lack of a clear separation between the low and high density quartiles does not contradict the results from the colour-density relation in Section~\ref{colour-mass-density}, where we possibly see evidence for a higher blue fraction in the most overdense environments above $\log~(1+\delta_3) \sim 0.8$. The high density quartile includes all galaxies at densities above $\log~(1+\delta_3) \sim 0.2$ and does not comprise the extreme overdensities only, in which we see the higher blue fraction. Since there is no relation with local density at densities below $\log~(1+\delta_3) \sim 0.8$, we do not see any difference between the colours of galaxies in the low and high quartile of the density distribution.

We conclude that in the redshift range of $1.5 \leq z \leq 3$ the highest local overdensities ($\log~(1+\delta_3) > 0.8$, i.e., an overdensity of roughly a factor of 5) possibly show a higher fraction of blue galaxies. At densities below $\log~(1+\delta_3) \sim 0.8$ there is no correlation between rest-frame colours or fraction of blue galaxies with local density. This finding is different from the colour density relation found at lower redshifts (up to $z\sim1.4$), where higher density environments show a higher fraction of red galaxies \citep[e.g.][]{Coo06,Cuc06,Cas07,Pan09,Tas09,Iov10,Gru10}.

\begin{figure*}
\includegraphics[width=0.485\textwidth]{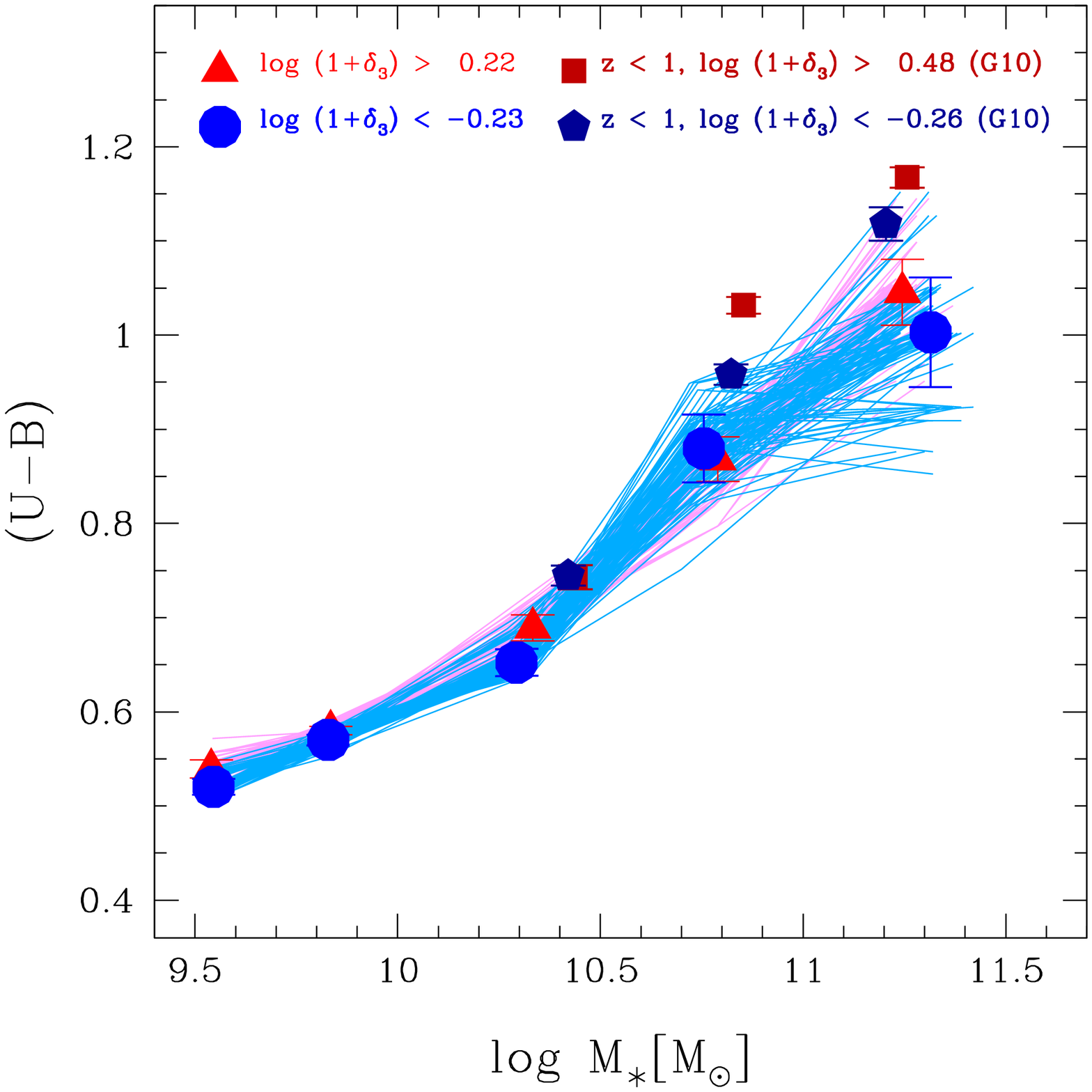}
\includegraphics[width=0.485\textwidth]{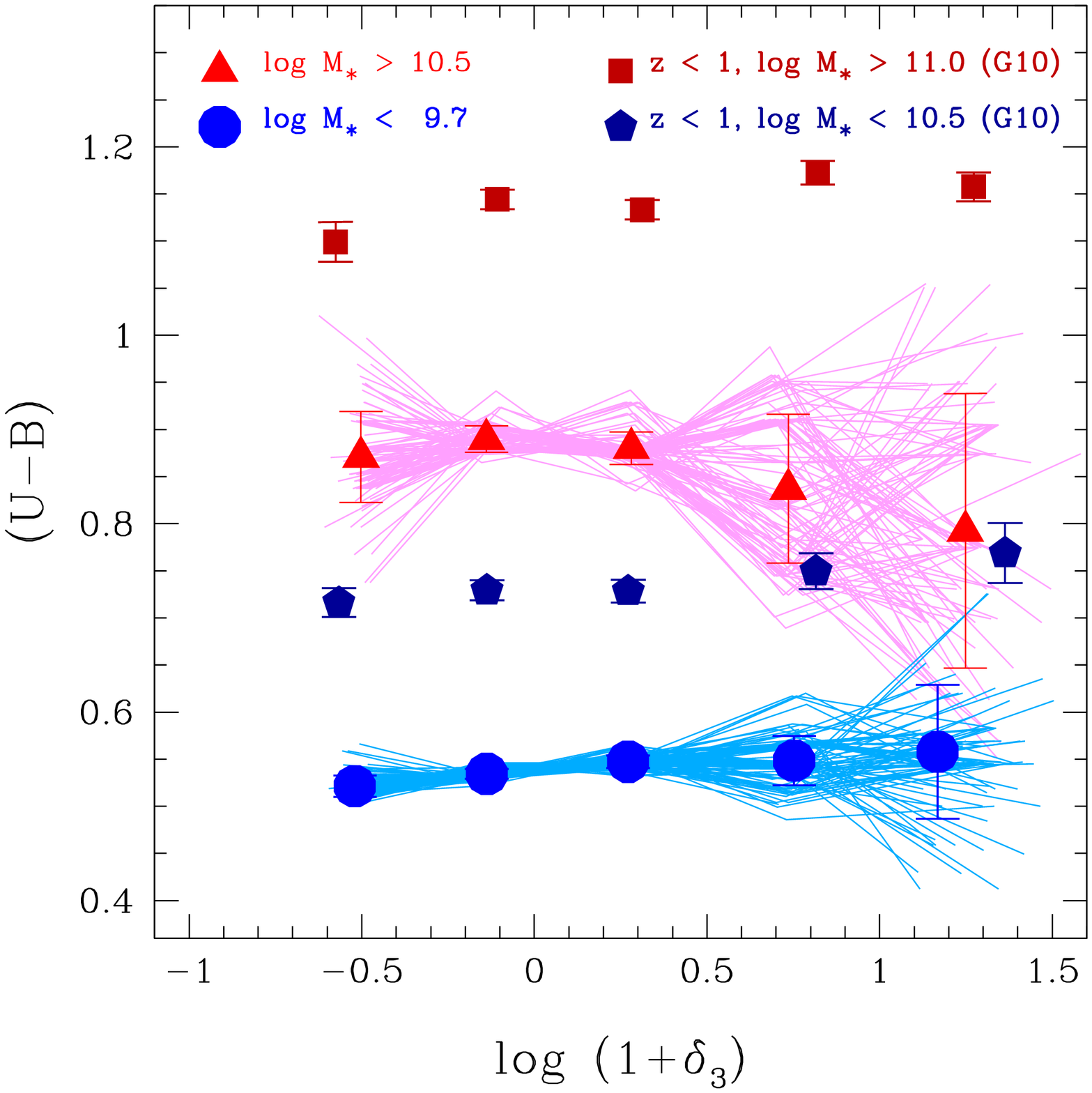}
\caption{Colour-stellar mass and colour density relation in low and high density and mass quartiles, respectively, over the whole redshift range $1.5 < z <3$. Left panel: colour-stellar mass relation in high (red) and low (blue) local density quartiles. Right panel: colour-density relation in high (red) and low (blue) stellar mass quartiles. This figure shows the results of the Monte Carlo simulations (see text). The mean and RMS of all Monte Carlo runs in bins of stellar mass (left panel) and local density (right panel) are overplotted as big symbols: red triangles for the high quartile and blue circles for the low quartile. The average data points from the sample of \citet[][(G10)]{Gru10} covering the redshift range $0.4 < z < 1$ are plotted as big red squares (high quartile) and blue pentagons (low quartile) respectively. \label{fig10}}
\end{figure*}

Figure~\ref{fig9} shows two different local densities $\log~(1+\delta_3)$ and $\log~(1+\rho)$ and $(U-B)$ colour in the low and high quartiles of the stellar mass distribution. The high quartile includes galaxies with $\log~M_\ast > 10.5$ (red), while the low quartile corresponds to $\log~M_\ast < 9.7$ (blue).
There are strong differences in the rest-frame colour distributions of low- and high-mass galaxies at all redshifts. As already investigated in the previous section, high-mass galaxies have consistently redder colours ($\Delta (U-B) \sim 0.4$ magnitudes). These differences are highly significant. The colour difference of $\sim$0.4 mag between the high- and low-mass quartile appears constant with redshift up to $z\sim 2.8$. The magenta line in the colour panel of Figure~\ref{fig9} shows the distinction between red and blue galaxies calculated from the red sequence at $z\sim1$ evolved back in time assuming passive evolution of the stellar populations (see Section~\ref{completeness-colour}). It corresponds to the position of the blue line at $M_B=-21.5$ in the colour-magnitude diagrams in Figure~\ref{fig5}. 
At a redshift of $z\sim 2.8$ the median colour of the high mass quartile decreases slightly towards the region occupied by the blue cloud. However, a similar decrease is seen at redshifts below $z<1.5$ and is probably caused by the target selection of the GNS, which favours the inclusion of very massive red galaxies between $1.7 < z < 3$ in our sample. This could slightly bias the average colour of massive galaxies towards redder colours within this redshift range.

We see slight evidence for a mass segregation emerging at $z \sim 1.8$, where more massive galaxies tend to be located preferentially in high density areas (see Figure~\ref{fig9}). This tentative result is consistent with what is found at lower and intermediate redshifts, where more massive galaxies tend to be located in high density environments \citep[see e.g.][]{Gru10}.

\subsection{Is the colour-density relation caused by variations in stellar mass?} \label{highlow}

Recent studies at intermediate redshift ($z\sim1$, see e.g. \citet{Tas09,Iov10,Gru10}) have suggested that the observed colour-density relation is mainly caused by the strong correlation between rest-frame colour and stellar mass. This also requires the presence of a stellar mass-density relation, such that higher mass galaxies are preferentially located in regions of higher overdensity. We indeed find a weak trend for this behaviour, emerging at $z < 1.8$, at a significance of $\sim 2\sigma$ (see Figure~\ref{fig9}). This is roughly consistent with the extrapolation of a similar weak positive correlation between stellar mass and local density found at $z<1$ by \citet{Gru10}, which seems to get steeper and more significant with decreasing redshift.

Another question is if the higher fraction of blue galaxies in high overdensities (see Section~\ref{colour-mass-density}) is due to a trend in stellar mass. To answer this question we investigate the colour-stellar mass relation and the colour-density relation in quartiles of local density and stellar mass respectively.

Figure~\ref{fig10}, left panel, shows the colour-stellar mass relation in high and low density quartiles. The right panel of Figure~\ref{fig10} shows the colour-density relation in high and low stellar mass quartiles. The same high and low quartiles of local density and stellar mass are used as described above (Section~\ref{extreme}). As in Figures~\ref{fig8} and \ref{fig9} we plot the results of each Monte Carlo run as well as the average and RMS of all runs instead of the measured data. The whole redshift range of $1.5 < z < 3$ is shown in Figure~\ref{fig10}.

To compare our result with the colour-density and colour-stellar mass relations at lower redshift we use the data of \citet[][G10 hereafter]{Gru10}. This sample is based on a deep near-infrared survey (the POWIR survey, \citet{Con08a}) and spectroscopic redshifts from the DEEP2 redshift survey \citep{Dav03}. The local densities and colours are measured in a similar way as in the present study. We split the G10 sample in high and low quartiles of local density and stellar mass as described above. The data covers the redshift range  $0.4 < z <1$. The median $(U-B)$ colours of this sample are overplotted as big symbols: dark red boxes are used for the high quartile and dark blue pentagons for the low quartile of the POWIR sample data-points respectively.

The left panel of Figure~\ref{fig10} shows that the strong correlation between $(U-B)$ and $\log~M_\ast$ is present in both, high (red) and low (blue) density quartiles. The relation is very similar in both density quartiles.
The colour-stellar mass relation at lower redshifts of $0.4 < z < 1$ (squares and pentagons in Figure~\ref{fig10}) is remarkably similar to the colour-stellar mass relation in the GNS data (triangles and circles), with a similar slope and a slightly larger colour difference between low and high local density quartiles ($\Delta(U-B) \sim 0.15$ mag). At the high mass end low $z$ galaxies have redder average colours than galaxies of the same stellar mass at high $z$. The colours of $\log~M_\ast \sim 10.5$ galaxies at $1.5 < z <3$ are indistinguishable from the colours of similar mass galaxies at $0.4 < z <1$.

\begin{figure*}
\includegraphics[width=0.37\textwidth, angle=270]{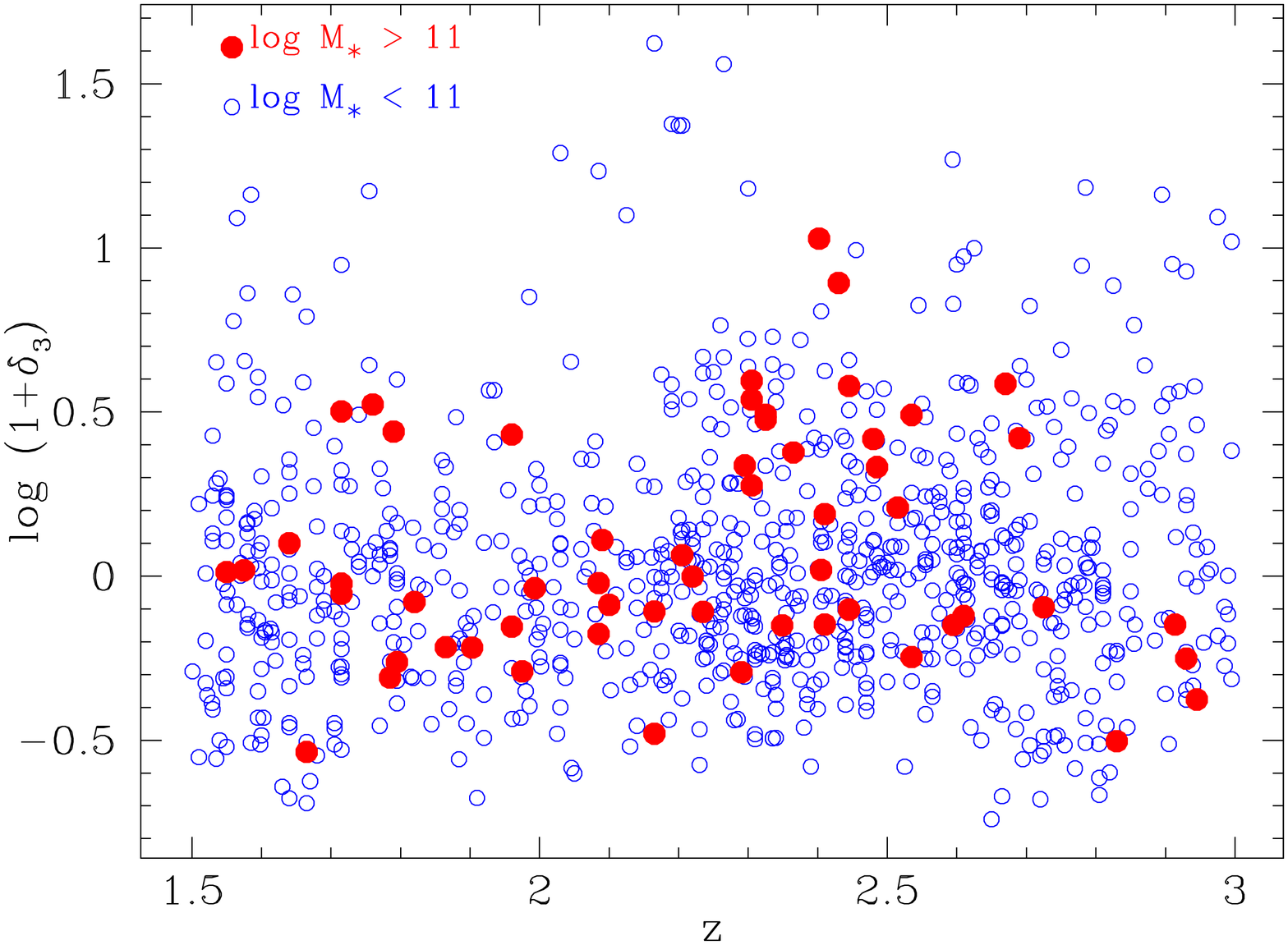}
\includegraphics[width=0.37\textwidth, angle=270]{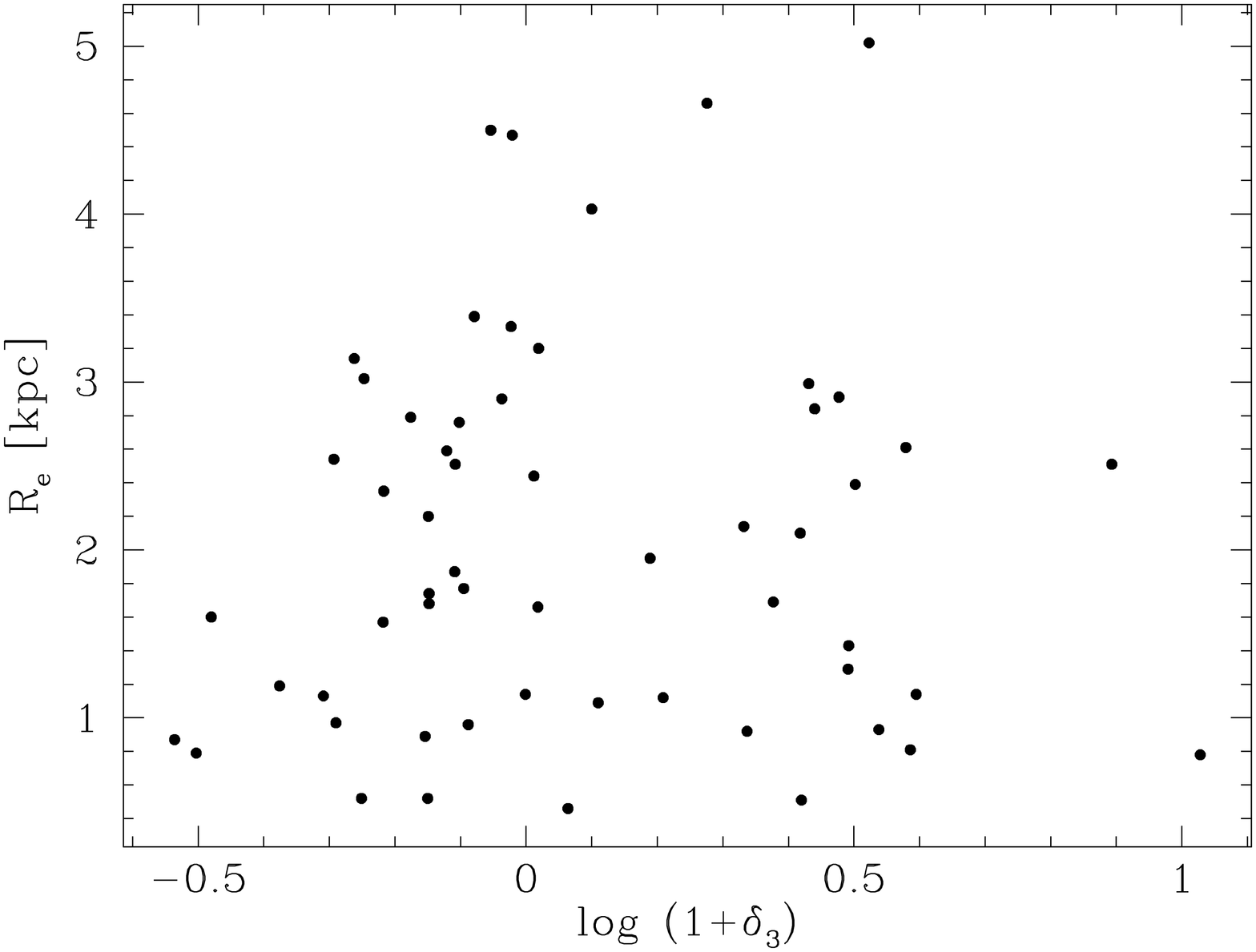}
\caption{Environments of most massive galaxies ($\log~M_\ast > 11$). Left panel: local density $\log~(1+\delta_3)$ for the full sample (blue open circles) and the most massive galaxies (red dots) as a function of redshift. Right panel: effective radii in kpc as a function of relative overdensity $\log~(1+\delta_3)$. \label{fig11}}
\end{figure*}

The right panel of Figure~\ref{fig10} shows the correlation between colour and local density for galaxies in the low and high stellar mass quartile. A clear colour offset between low- and high-mass galaxies of $\Delta(U-B) \sim 0.2-0.3$ mag is present at all densities. Interestingly, there is a trend that mean colours of galaxies in the high-mass quartile are bluer at higher overdensities ($\log~(1+\delta_3)>0.5$) than at low and average local densities. For galaxies in the low-mass quartile we do not see a strong correlation between colour and local density. 
Combined with the results of the colour-density relation in Figure~\ref{fig7}, this would point towards high-mass galaxies being responsible for the higher fraction of blue galaxies at the highest overdensities. Note however that the RMS scatter between the single Monte Carlo simulations is larger than the decrease in colour. This large spread is due to the low numbers of objects at the highest overdensities (33 galaxies at $\log~(1+\delta_3)>0.8$.
The colour difference between the GNS data-points and the $z<1$ data-points from G10 is mainly due to the differences in the stellar mass distribution in the two surveys, leading to different limits of the low and high quartiles. The GNS reaches much lower stellar masses down to $\log~M_\ast \sim 9.5$, whereas the completeness limit of the POWIR survey is $\log~M_\ast \sim 10.25$. Additionally, the G10 sample has a larger number of high mass galaxies with $\log~M_\ast>11$ than the GNS sample, leading to redder average colours, as expected from the colour-mass relation.

If the colour-density relation we see for high-mass galaxies is real, then it is reversed with respect to what is found at lower and intermediate redshifts up to $z\sim0.85$, where galaxies in higher local densities tend to be redder \citep[see e.g..][]{Kau04,Pat09,Gru10}. Figure~\ref{fig10} shows that the high and low mass quartiles of the G10 sample (up to $z\sim1$) show a weak correlation between colour and local density, such that the average colour increases with higher local density. On the other hand, it was argued by some studies that the colour-density relation and colour-SFR relation at $z\sim1$ might be reversed \citep{Elb07,Coo08}. Note, however, that \citet{Pat09} argue that the reversed SFR-density relation (were galaxies in denser environments have higher SFRs) is mainly driven by starforming {\it low}-mass galaxies, whereas our data indicate that the reversed colour-density relation is due to on average bluer {\it high}-mass galaxies. In other worlds, we see a lack of red high-mass galaxies in regions of high relative overdensities. The on average bluer high-mass galaxies, however, are on average still redder than low-mass galaxies at comparable densities, but bluer than their high-mass counterparts at average and low local densities. 
Note that the bluer colour of high mass galaxies does not necessarily imply that there is more ongoing star formation in these galaxies, but that there is possibly {\it less dust-attenuated star formation} at highest overdensities. Massive galaxies in our sample show an overall very high dust content relative to low mass galaxies (Bauer et al. submitted). The behaviour of the SFR-density relation using dust corrected star formation rates from Bauer et al. (submitted) will be investigated in a forthcoming paper.

Recently, a study of the colour-density relation at redshifts up to $z\sim2.7$ was performed by Chuter et al. (submitted) using the UKIDSS (UKIRT Infrared Deep Sky Survey) Ultra Deep Survey (UDS) data. They find that the local colour-density relation, with redder galaxies located in regions of higher local density, is present up to a redshift of $z\sim1.75$. Above this redshift the distinction between the environments of red and blue galaxies can not be confirmed at a statistically significant level. The authors also report a possible reversal of the colour-density relation at even higher redshift, however, the currently available UDS data release (DR3) does not allow for reliable conclusions at high $z$. The results we present in this study are consistent with extrapolating the results of Chuter et al. (submitted), suggesting a gradual disappearance of the colour density relation with redshift and a possible reversal of the colour-density relation at the most extreme overdensities at $z>1.5$.

\subsection{The environment of the most massive galaxies and its relation to galaxy sizes}

In the following we investigate how, and if, the environments of the most massive galaxies ($\log~M_\ast > 11$) differ from the rest of the galaxy population and if there is a correlation between local density and galaxy size. The subsample of massive galaxies for which a size measurement is available comprises 57 galaxies from the selection of massive galaxies in the GOODS fields on which the GNS pointings were centred \citep{Con10}. 
The measured local densities of those galaxies are very reliable since they are in the centre of their respective pointing and should not be affected by survey edges, as discussed in Section~\ref{density}. The left panel of Figure~\ref{fig11} plots the local densities of the massive galaxy sample (red dots) and the rest of the sample (blue circles) against redshift. A K-S test between the two subsamples shows that they are not statistically different.

To examine a possible environmental effect on the sizes of massive galaxies in our sample, we use the galaxies' effective radii measured by \citet{Bui08}. They argue that the measured effective radii of these massive galaxies are consistently smaller than the typical effective radii of galaxies of comparable stellar mass in the local universe. Different scenarios have been proposed to account for this size evolution, one of them being dry merging. In this scenario repeated minor merging of gas poor galaxies would not trigger the formation of new stars, but could possibly increase the size of a galaxy by dynamical friction and the injection of angular momentum \citep{Naa09}. This scenario would then suggest a connection between galaxy sizes and local density, since it requires the presence of numerous companions for galaxies to grow in size.
The right panel of Figure~\ref{fig11} shows the effective radii $R_{e}$ of massive galaxies as a function of their local density $\log~(1+\delta_3)$. A Spearman Rank correlation test finds no significant correlation between $R_{e}$ and local density, although all galaxies in the most underdense environments ($\log~(1+\delta_3)<-0.2$) have small $R_{e}$ ($<$ 3 kpc). However, due to the small number of objects in our sample this effect probably arises by chance.

The lack of a correlation between galaxy size and local density is probably due to the different effects that the environment might have on galaxies. High local densities may not only increase the size of a galaxy via minor mergers as suggested by e.g. \citet{Naa09}. High density environments might as well tidally truncate galaxies \citep[see e.g.][]{Rub88}, although tidal truncation is most effective on low-mass galaxies. Another possibility is that the size evolution, does not occur through frequent minor merging of satellite galaxies, but through internal processes like AGN feedback \citep{Fan08}. 

\section{Summary and conclusions}

In this study we investigate the influence of stellar mass and local density on galaxy rest-frame colour and the fraction of blue galaxies at redshifts between $1.5<z<3$ based on observational data from a deep HST $H$-band survey of unprecedented depth, the GOODS NICMOS Survey (GNS), reaching a stellar mass completeness limit of $M_\ast=10^{9.5}$ at $z=3$. We find the following results:

\begin{enumerate}
\renewcommand{\theenumi}{\arabic{enumi}.}

\item Galaxy colour depends strongly on galaxy stellar mass at all redshifts up to $z\sim3$. The colour-stellar mass relation does not evolve with redshift below $z\sim3$. The stellar mass where the blue and red fractions cross over is roughly constant at $\log~M_\ast = 10.8$ in the redshift range between $1.5 \leq z \leq 3$, which is remarkably similar to the cross-over mass found at lower redshifts between $0.4 < z < 1$.\\ 

\item The strong colour-stellar mass relation is very similar across all local environments and does not evolve strongly with redshift. The colour-stellar mass relation at $1.5 < z < 3$ has a slope similar to the relation at lower redshift ($0.4 < z < 1$), which also has a larger offset in colour between low and high local densities. At the same stellar mass, massive galaxies at high redshifts ($1.5 < z < 3$) are bluer than at low redshifts ($0.4 < z < 1$), whereas the average colour of lower mass galaxies ($\log~M_\ast \sim 10.5$) does not vary strongly with redshift.\\

\item We do not find a strong influence of local environment on galaxy colours. If the colour-density relation persists at $z>1.5$ it must be very weak. We determine an upper limit to the  possible change in the average $(U-B)$ colour between high and low relative densities of $\sim 0.1$ magnitudes. However, the most overdense regions ($\sim 5$ times overdense) may be populated by a higher fraction of blue galaxies than average and underdense regions. The difference in $f_{blue}$ is $\sim 10\%$. We find possible evidence that this higher blue fractions at the most extreme overdensities could be caused by a lack of massive red galaxies at the highest local densities. \\

\item We do not find a significant correlation between galaxy sizes (effective radii) and relative overdensity, although we do not find any galaxies with large effective radii ($R_e > 3$kpc) at very low densities ($\log~(1+\delta_3)<-0.2)$. However, this might be due to the small number of objects we find in low density environments and is probably a chance effect.\\

\end{enumerate}

To summarise, our data suggests that stellar mass is the most important factor in determining the colours of galaxies in the early universe up to $z\sim3$. Local density might have a small additional effect, but only at the most extreme overdensities, which are populated by a higher fraction of blue galaxies. These results are consistent with studies at lower and intermediate redshift that suggest a gradual weakening of the environmental influence with higher redshift. A possible interpretation for this is that the environmental processes that alter the properties of galaxies are proceeding slowly over cosmic time. Some of the most influential high density environments like galaxy clusters are still in the process of being build-up at $z>1.5$ and yet cannot change galaxy colours via e.g., ram pressure stripping or strangulation.

If the trend for higher blue fractions at the highest local densities is real, it would suggest that we are witnessing the epoch of high star formation in more massive galaxies and that the local environment contributes to this epoch by triggering SF through galaxy interactions. Note however, that the bluer colours at high local densities could as well be due to lower dust attenuation in these environments. The star formation rates of galaxies in the GNS are currently investigated by Bauer et al. (submitted) and will be the focus of a subsequent study of the relation between SFR, stellar mass and local density (Gr\"utzbauch et al., in preparation).
Future deep and wide surveys such as the UKIDSS UDS and VISTA will be better able to address the environment vs. stellar mass issue in more detail in the coming years.

\section*{Acknowledgments}

RG would like to thank Jes\'us Varela for many useful discussions and very helpful suggestions.
We acknowledge funding from the UK Science and Technology Facilities Council (STFC).
We also thank the anonymous referee for their suggestions, which greatly improved the content of this paper.

\end{document}